\documentclass[useAMS,usenatbib,graphicx]{mn2e}
\usepackage[hyphens]{url}
\usepackage{times}
\usepackage{epsfig}
\usepackage{amssymb}

\title[Multicolour modelling of SN~2013dx associated with GRB~130702A.]{Multicolour modelling of SN~2013dx associated with GRB~130702A$^\ast$}

\author[A. A. Volnova et. al.]{
A. A.~Volnova,$^{1}$\thanks{email:~alinusss@gmail.com} M.~V.~Pruzhinskaya,$^{2,3}$\thanks{email:~pruzhinskaya@gmail.com}, A.~S.~Pozanenko,$^{1,4,5}$ S.~I.~Blinnikov,$^{6,7,8}$\\
\newauthor P.~Yu.~Minaev,$^{1}$ O.~A.~Burkhonov,$^{9}$ A.~M.~Chernenko,$^{1}$ Sh.~A.~Ehgamberdiev,$^{9}$ R.~Inasaridze,$^{10}$\\
\newauthor M.~Jelinek,$^{11}$ G.~A.~Khorunzhev,$^{1}$ E.~V.~Klunko,$^{12}$ Yu.~N.~Krugly,$^{13}$ E.~D.~Mazaeva,$^{1}$\\
\newauthor V.~V.~Rumyantsev,$^{14}$ A.~E.~Volvach$^{14}$\\
$^{1}$Space Research Institute, 84/32 Profsoyuznaya Street, Moscow 117997, Russia\\
$^2$Lomonosov Moscow State University, Sternberg Astronomical Institute, Universitetsky pr., 13, Moscow, 119234, Russia\\
$^3$Laboratoire de Physique Corpusculaire, Universit\'e Clermont Auvergne, Universit\'e Blaise Pascal, CNRS/IN2P3, Clermont-Ferrand, France\\
$^4$National Research Nuclear University MEPhI (Moscow Engineering Physics Institute), 115409 Moscow, Russia\\
$^5$Moscow Institute of Physics and Technology, 9 Institutskiy per., Dolgoprudny, Moscow Region, 141700, Russia\\
$^6$Institute for Theoretical and Experimental Physics, Bolshaya Cheremushkinskaya ulitsa 25, 117218 Moscow, Russia\\
$^7$All-Russia Research Institute of Automatics, Sushchevskaya ulitsa 22, 127055 Moscow, Russia\\
$^8$Kavli Institute for the Physics and Mathematics of the Universe (WPI), The University of Tokyo Institutes for Advanced Study, The University of Tokyo, \\5-1-5 Kashiwanoha, Kashiwa, 277-8583, Japan\\
$^9$Ulugh Beg Astronomical Institute (UBAI) of the Uzbek Academy of Sciences, 33 Astronomicheskaya str., 
Tashkent, 100052, Uzbekistan\\
$^10$Kharadze Abastumani Astrophysical Observatory, Ilia State University, Kakutsa Cholokashvili Ave 3/5
Tbilisi 0162, Georgia\\
$^{11}$Astronomical Institute of the Czech Academy of Sciences, Fri\v{c}ova 298, 251 65 Ond\v{r}ejov, Czech Republic\\
$^{12}$Institute of Solar-Terrestrial Physics, Russian Academy of Sciences, Siderian branch, 664033, Irkutsk p/o box 291; Lermontov st., 126a, Russia\\
$^{13}$Kharkiv National University, Institute of Astronomy, 35 Sumska Str., Kharkiv, 61022, Ukraine\\
$^{14}$Crimean Astrophysical Observatory of the Russian Academy of Sciences, 298409, Crimea, Bakhchisaray region, Nauchny, Russia\\
$\ast$Based on observations made with the Nordic Optical Telescope, operated on the island of La Palma jointly by Denmark, Finland, Iceland, Norway, \\
and Sweden, in the Spanish Observatorio del Roque de los Muchachos of the Instituto de Astrofisica de Canarias.
}
\begin{document}

\date{Accepted XXXX. Received XXXX.}
\pagerange{\pageref{firstpage}--\pageref{lastpage}} \pubyear{XXXX}
\label{firstpage}
\maketitle

\begin{abstract} 
We present optical observations of SN~2013dx, related to the Fermi burst GRB~130702A occurred at a redshift $z=0.145$. It is the second-best sampled GRB-SN after SN~1998bw: the observational light curves contain more than 280 data points in $uBgrRiz$ filters until 88 day after the burst, and the data were collected from our observational collaboration (Maidanak Observatory, Abastumani Observatory, Crimean Astrophysical Observatory, Mondy Observatory, National Observatory of Turkey, Observatorio del Roque de los Muchachos) and from the literature. We model numerically the multicolour light curves using the one-dimensional radiation hydrodynamical code \textsc{stella}, previously widely implemented for the modelling of typical non-GRB SNe. The best-fitted model has the following parameters: pre-supernova star mass $M = 25~\rm M_{\sun}$, mass of a compact remnant $M_{\rm CR} = 6~\rm M_{\sun}$, total energy of the outburst $E_{\rm oburst} = 3.5 \times 10^{52}$ erg, pre-supernova star radius $R = 100~\rm R_{\sun}$, $M_{\rm^{56}Ni} = 0.2~\rm M_{\sun}$ which is totally mixed through the ejecta; $M_{\rm O} = 16.6~\rm M_{\sun}$, $M_{\rm Si} = 1.2~\rm M_{\sun}$, and $M_{\rm Fe} = 1.2~\rm M_{\sun}$, and the radiative efficiency of the SN is $0.1$ per cent.
\end{abstract}


\section{Introduction}

The observational association between long-duration gamma-ray bursts (GRBs) and Type Ib/c supernovae (SNe) has been confirmed during last two decades, supporting the connection between GRBs and the death of massive stars 
\citep[see, e.g., ][]{hjorthbloom}. The first reliable association between a GRB and a SN was that of GRB~980425 with SN~1998bw, at a redshift $z = 0.0085$ \citep{galama,iwamoto,kulkarni}. Since then about 40 supernovae associated with GRBs were discovered, and a half of them are spectroscopically confirmed (e.g., SN~2003dh, \citealp{2003dh}; SN~2006aj, \citealp{2006aj}; SN~2010bh, \citealp{2010bh}; SN~2013cq, \citealp{2013cq}; SN~2013fu, \citealp{2013fu}, \citealp{cano16}).

To determine the physical properties of the SN explosion and/or of its progenitor (a mass of $\rm^{56}$Ni, an ejecta mass $M_{\rm ej}$, a total energy of the explosion $E_{\rm oburst}$ etc.) detailed photometrical and spectroscopic data are necessary. Due to the lack of observations the most common way to estimate the physical parameters of GRB-SNe is to use some ``classical'' light curves and spectra of well-studied SNe as templates \citep[e.g., ][]{cano-model}. Most often the multicolour and bolometric light curves of SN~1998bw and 2003dh are used. However, this empirical method of modelling is based on rather simple assumptions about SN explosions, initial conditions and evolution \citep{arnett}.

In this paper we are modelling the multicolour light curves of SN~2013dx, associated with GRB~130702A, using the code \textsc{stella}~\citep{Blinnikov98,Blinnikov06}. \textsc{stella} is a package of one-dimensional spherically symmetrical multi-group radiation hydrodynamics code which treats non-equilibrium radiative transfer according to chemical composition and inner structure of a pre-supernova star. The code has been used for light curve modelling of different types of SNe (Ia,~\citealp{Blinnikov06}; Ib/Ic,~\citealp{Folatelli2006,Tauris2013}; IIb,~\citealp{Blinnikov98}; IIn,~\citealp{Chugai2004}; IIP,~\citealp{Baklanov2005,Tominaga2009}). The assumptions about the supernova outburst geometry are also simple like in the empirical method, but the consideration of chemical abundances and distribution of different chemical elements inside a pre-supernova star allows one to calculate radiative transfer during the explosion and to build more physically correct modelled light curve. 

With the \textsc{stella} code one can model many properties
of the SN explosion. Spectra calculated for every specific time
since the explosion allow one to model the multicolour light curves and
photospheric velocities of the SN expanding envelope. The calculations
take into account distribution of abundance of chemical elements
in the envelope before the explosion, the interaction between the
inner layers and the compact core, and so on. Another advantage
of modelling is that it makes it possible to investigate a line-of-sight extinction by the circumburst medium by comparing modelled and observed light curves in various photometric filters.

The results from the \textsc{stella} code were found to be in a good agreement with those from other well-known hydro-dynamic codes \citep[e.g., ][]{Sedona2,Artis,Sedona,Kozyreva2017}.

In Section~\ref{sec:obs} we present the observational data used for the modelling: we construct the detailed multicolour light curves using original observations and published data. In Section~\ref{sec:lc} we explain the process of SN light curve extraction. In Section~\ref{sec:modelling} we describe the modelling procedure with the \textsc{stella} code and present the properties of the model which best matched to the observed light curves of SN~2013dx. In Section~\ref{sec:discussion} we discuss remaining unresolved questions about the properties of both GRB~130702A and SN~2013x.

\section{Observations}
\label{sec:obs}
\subsection{Detection of GRB~130702A / SN 2013dx}
\label{sec:obsgamma}

GRB~130702A (= Fermi trigger 394416326) triggered the Gamma-ray Burst Monitor \citep[GBM;][]{GBM} and was observed by the Large Area Telescope \citep[LAT;][]{LAT} aboard the space observatory \textit{Fermi} 
at 00:05:23.079 \textsc{ut} on July 2, 2013 \citep{cheung,collazzi}. Hereafter we consider
this time as $T_0$, and all time intervals since the burst trigger are referred as $t$. The burst has duration $T_{90}$ \citep[the time during which the cumulative
counts increase from 5\% to 95\% above background; ][]{Kouveliotou} of about 59~s in the GBM 
energy range of 50--300 keV. \textit{Fermi}/LAT detected more than 5 photons above 100
MeV up to $t = 2200$ s, and the highest energy photon is a
1.5 GeV event which was observed 260 s after the GBM trigger \citep{cheung}. 
The best LAT location found by \citet{cheung} during on-ground analysis is: $\alpha =
216\fdg4$, $\delta = +15\fdg8$ (J2000), with an error radius of $0\fdg5$ (90 per cent containment, statistical
error only). This position is in $4\degr$ from the best GBM position ($\alpha = 218\fdg81$, $\delta = +12\fdg25$ with 1$sigma$ uncertainty of $4\degr$), and in $0\fdg8$ from the position of the optical afterglow (OA)~\citep{singer-gcn}.

The OA of GRB~130702A was discovered by the intermediate Palomar Transient Factory
\citep{PTF,iPTF} with the Palomar 48-inch Oschin telescope (P48) \citep{singer-gcn,singer_paper}. 
The source was labelled as iPTF13bxl with coordinates $\rm\alpha = 14^h29^m14\fs78$, $\rm\delta = +15\degr46\arcmin26\farcs4$
(J2000). The OA is located in the vicinity of two SDSS sources: a bright galaxy SDSS J142914.57+154619.3, 
at a separation of $7\farcs6$ and a faint source SDSS J142914.75+154626.0 classified as a star in the catalogue, 
at a separation of $0\farcs6$. The latter was suggested to be a galaxy rather than a star, and to be the host galaxy of the burst in subsequent 
studies (\cite{kelly};~\cite{delia}, hereafter D15;~\cite{toy}, hereafter T16). 

The redshift of the OA was obtained based on the detection of H $\alpha$, O \textsc{ii}, and O \textsc{iii} emission corresponding to $z = 0.145$ \citep{z1,z2,z3}. The redshift is consistent with that of one of the nearby bright galaxies SDSS J142914.57+154619.3 measured spectroscopically by \citet{leloudas}. This allowed \citet{kelly} to suggest that the host galaxy of GRB~130702A may be a dwarf satellite of an adjacent massive spiral galaxy.

The emerging supernova SN 2013dx associated with GRB~130702A was discovered photometrically about 6 days
after the burst trigger with the 2.5-m Nordic Optical Telescope (NOT) based on obvious brightening
of the OA and colour evolution unexpected for decaying afterglow \citep{schulze}. Later the supernova was confirmed
spectroscopically \citep{cenko,delia-gcn} and got the name SN 2013dx.

\subsection{Observations in gamma-rays}
\label{sec:gamma}

\begin{figure}
\centering
\includegraphics[width=90mm]{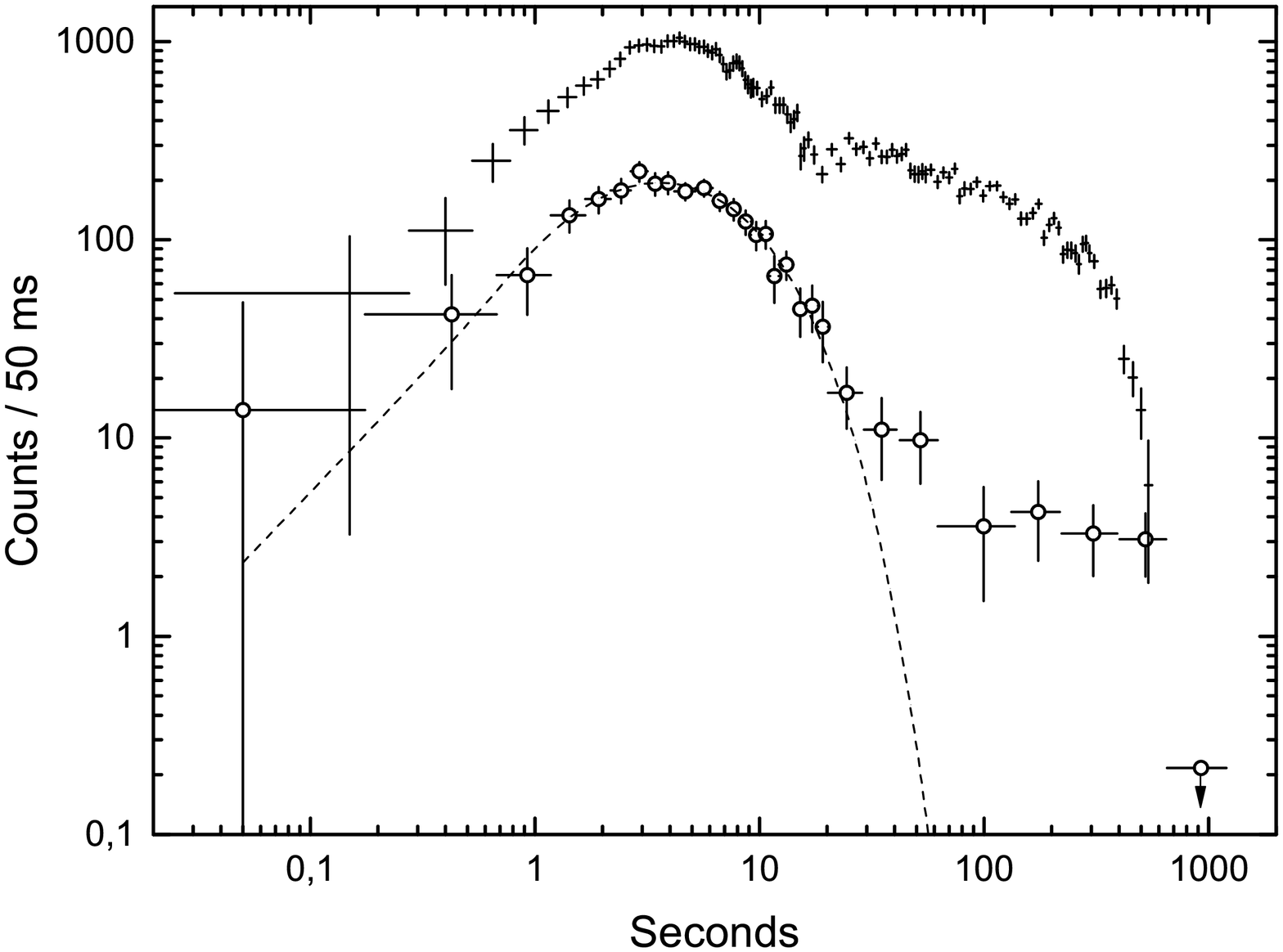}
\caption{Background subtracted light curve of GRB 130702A of SPI-ACS/\textit{INTEGRAL} (open circles, energy range 80--10000] keV) and GBM/\textit{Fermi} (crosses, energy range 9--900 keV). Dashed line represents the fit of main emission component by the exponential model \citep{norris}. Time in seconds since burst prompt phase is presented. The flux in counts is presented per 50 ms time interval. GBM/\textit{Fermi} light curve is multiplied by a factor of 10 for clarity. The phase of extended emission and sharp cut-off in its end are clearly visible in both GBM and SPI-ACS light curves.}
\label{spiacs}
\end{figure}

\begin{figure}
\centering
\includegraphics[width=90mm]{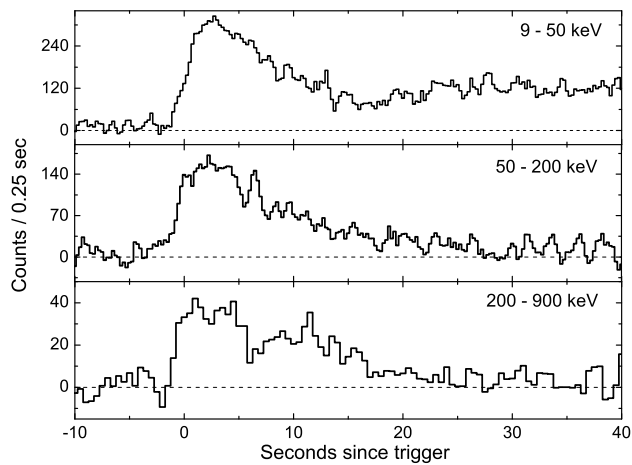}
\caption{Background subtracted light curve of GRB~130702A in three energy bands (GBM/\textit{Fermi}). Time resolution is 0.25 s for 9--50 keV and 50--200 keV energy bands and 0.5 s for 200--900 keV energy band. The horizontal axis is time since the GBM trigger, while the vertical axis is the gamma-ray flux of every energy band.}
\label{gbm}
\end{figure}

The burst was also observed by Konus-\textit{Wind} 
and had a fluence of $6.70_{-0.80}^{+0.82} \times 10^{-6}$ erg~cm$^{-2}$ in the 20--1200 
keV energy range \citep{konus_gcn}. The gamma-ray emission of the burst was also
detected by SPI-ACS/\textit{INTEGRAL} and GRNS/\textit{MESSENGER} \citep{integral}. 

We analysed GRB 130702A's prompt emission in the gamma-ray domain using
GBM/\textit{Fermi} and SPI-ACS/\textit{INTEGRAL} data. The light curve reveals two components in
both SPI-ACS and GBM data. These components are the main episode of emission
with duration of approximately 20 s and the extended emission with
duration up to 650 s (Fig.~\ref{spiacs}). The duration of the burst was calculated
in the energy range $>80$ keV using SPI-ACS data and found to be $T_{90} = 545 \pm 60 $ s.
An off-axis angle in SPI-ACS of the GRB 130702A is $69\degr$ and is close to
an optimal for detection.

The light curve of the main emission component in three energy channels of GBM is
presented in Figure~\ref{gbm}. In the soft energy channel, 9--50 keV, the time profile has a smooth FRED-like\footnote{FRED stands for Fast Rise and Exponential Decay -- the common shape of GRB pulses.} shape. The initial rising part of the second extended
emission component is also visible, starting at approximately 15 s after
the trigger. The extended emission in a hard energy channel is much weaker but it is also detectable up to
650~s after the trigger.

The spectral lag analysis was performed on GBM/\textit{Fermi} data for the main episode of emission in the time interval from -5 to 20 s from the trigger. The lag was calculated using CCF method
\citep{band} between the channels of 25--50 and 50--100 keV, and also between the channels of 25--50 and 100--300 keV. In both cases the spectral lag is almost negligible 
($0.3 \pm 0.3$ and $0.5 \pm 0.3$ s, respectively). 
According to \citet{norris2002} 80 per cent of long BATSE GRBs have the lag values less than 0.5 s between the channels of 25--50 and 100--300 keV.

Spectral analysis was performed for the main episode in the time interval from -1.5 to 13 s after the trigger using \texttt{rmfit}\footnote{\url{http://fermi.gsfc.nasa.gov/ssc/data/analysis/rmfit/}} 
and data from the BGO$_{01}$, NaI$_{06}$, NaI$_{07}$ and NaI$_{08}$ detectors of GBM/\textit{Fermi}. 
The energy spectrum is well fitted by a single
power-law model with index of $\gamma = -1.78 \pm 0.02$. The fluence of the
main emission episode in 10--1000 keV energy range is $(6.17 \pm 0.22) \times
10^{-6}$ erg~cm$^{-2}$.  

\subsection{\textit{Swift}-XRT observations}
\label{sec:xrt}

\begin{figure}
\centering
\includegraphics[width=90mm]{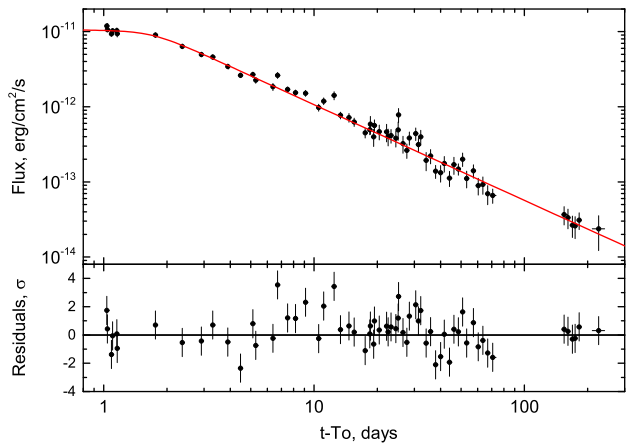}
\caption{\textit{Upper panel}: the X-ray light curve in the range of 0.3--10 keV, obtained by \textit{Swift}/XRT (black circles). The data before $t = 20$ d were binned for better signal-to-noise ratio. 
The solid line is a BPL fit with the jet-break time $t_{\rm jb} = 1.7 \pm 0.2$ d, and the slope after the break $\alpha_{X,2} = -1.27 \pm 0.02$.
\textit{Lower panel}: the residuals of the fit.}
\label{xrt}
\end{figure}

The X-ray counterpart of GRB~130702A was observed by the X-Ray Telescope \citep[XRT;][]{xrt}
aboard the \textit{Swift} space observatory \citep{swift} starting from 89.1 ks after the 
GBM trigger, i.e. $t = 1.03$ d~\citep{xrtobs}. The observations continued up to 225 d after 
the trigger\footnote{\url{http://www.swift.ac.uk/xrt_curves/00032876/ }}. 

We downloaded the full X-ray
light curve from the XRT Light Curve Repository \citep{evans} and fitted it with a broken power law \citep[BPL; see, e.g.,][]{beuermann} the using non-linear least squares method by \citet{Markwardt}. The best fit parameters are the following: slope before the break,  $\alpha_{X,1} = 0.03 \pm 0.26$, slope after the break, $\alpha_{X,2} = -1.27 \pm 0.02$ (Fig.~\ref{xrt}, upper panel), jet-break time, $t_{\rm jb} = 1.7 \pm 0.2$ d after the burst trigger, and sharpness of the break is fixed on 5, giving $\chi^2$/DOF = 72/58. These parameters differ from those obtained by \citet{singer_paper} because they used a restricted data set (the first 10 d instead of 225 d after the burst trigger) and fitted it with a single power law. We will assume that the afterglow light curve is achromatic after $t_b$ between X-ray and optical band-passes and use our best fit X-ray light curve parameters to subtract the afterglow trend from the optical light curve of GRB 130702A.

The results of our fit and the known redshift allow us to estimate the isotropic-equivalent energy release in the X-rays by integrating the obtained BPL function from 1~d to 225~d after the trigger:
$E_{\rm iso,\it X} > 2.7 \times 10^{50}$ erg, and it is a lower limit because of the absence 
of X-ray data before $t = 1.03$~d.

\subsection{Optical follow-up}

\subsubsection{Our observations}

\begin{table*}
\centering
 \begin{minipage}{140mm}
  \caption{Log of our observations. All magnitudes are in AB system and not corrected for Galactic extinction.}
  \begin{tabular}{c|c|c|c|c|c}
  \hline
Date & \textsc{ut} start & $T_0+$ & Telescope & Filter & Magnitude\\
  \hline
2013-07-03 & 17:38:18 & 1.7083 & AZT-22 & $R$     & $18.88 \pm 0.04$ \\
2013-07-03 & 17:40:00 & 1.7554 & AS-32  & Clear & $18.89 \pm 0.03$\\
2013-07-04 & 16:49:20 & 2.6743 & AZT-22 & $R$     & $19.43 \pm 0.04$\\
2013-07-04 & 19:32:52 & 2.7864 & AS-32  & Clear & $19.33 \pm 0.07$\\
2013-07-05 & 16:25:11 & 3.6592 & AZT-22 & $R$     & $19.79 \pm 0.02$\\
2013-07-05 & 20:23:21 & 3.8377 & ZTSh   & $R$     & $19.86 \pm 0.02$\\
2013-07-06 & 16:20:12 & 4.6540 & AZT-22 & $R$     & $19.89 \pm 0.03$\\
2013-07-07 & 16:45:34 & 5.6716 & AZT-22 & $R$     & $20.01 \pm 0.04$\\
2013-07-08 & 16:49:14 & 6.6759 & AZT-22 & $R$     & $19.99 \pm 0.05$\\
2013-07-09 & 16:50:52 & 7.6770 & AZT-22 & $R$     & $19.91 \pm 0.04$\\
2013-07-10 & 17:31:29 & 8.6931 & AZT-22 & $R$     & $19.83 \pm 0.03$\\
2013-07-11 & 16:50:14 & 9.6644 & AZT-22 & $R$     & $19.80 \pm 0.03$\\
2013-07-12 & 17:19:13 & 10.6846 & AZT-22 & $R$    & $19.78 \pm 0.03$\\
2013-07-13 & 20:26:12 & 11.8144 & Z-1000/CrAO & $R$    & $19.71 \pm 0.04$\\
2013-07-14 & 17:08:02 & 12.6768 & AZT-22 & $R$    &  $19.70 \pm 0.04$\\
2013-07-15 & 16:44:09 & 13.6602 & AZT-22 & $R$     & $19.69 \pm 0.04$\\
2013-07-15 & 19:23:12 & 13.7853 & AS-32  & Clear & $19.65 \pm 0.06$\\
2013-07-15 & 19:41:31 & 13.7988 & Z-1000/CrAO & $R$     & $19.69 \pm 0.04$\\
2013-07-16 & 17:25:47 & 14.6891 & AZT-22 & $R$     & $19.65 \pm 0.05$\\
2013-07-16 & 21:08:17 & 14.8524 & Z-1000/CrAO & $R$     & $19.68 \pm 0.07$\\
2013-07-16 & 21:30:57 & 14.8678 & AS-32  & Clear & $19.58 \pm 0.07$\\
2013-07-18 & 16:56:54 & 16.6691 & AZT-22 & $R$     & $19.62 \pm 0.04$\\
2013-07-19 & 16:54:58 & 17.6677 & AZT-22 & $R$     & $19.74 \pm 0.04$\\
2013-07-21 & 17:29:40 & 19.6918 & AZT-22 & $R$     & $19.72 \pm 0.05$\\
2013-07-23 & 17:23:24 & 21.6875 & AZT-22 & $R$  &    $19.75 \pm 0.05$\\
2013-08-03 & 19:40:37 & 32.7930 & ZTSh & $R$    &  $20.70 \pm 0.03$\\
2013-08-04 & 18:55:51 & 33.7762 & ZTSh & $R$    &  $20.71 \pm 0.03$\\
2013-08-15 & 11:43:16 & 44.4624 & AZT-22 & $R$     & $21.13 \pm 0.12$\\
2013-08-18 & 11:38:02 & 47.4688 & AZT-22 & $R$     & $21.44 \pm 0.11$\\
2013-08-28 & 11:48:54 & 57.4622 & AZT-22 & $R$     & $21.77 \pm 0.07$\\
2013-08-28 & 18:39:39 & 57.7404 & RTT-150 & $R$    & $21.83 \pm 0.14$\\
2013-09-27/28 & 20:02:28 & 88.3032 & NOT   & $R$    & $22.40 \pm 0.09$\\
2014-03-30 & 18:48:51 & 271.7885 & AZT-33IK & $R$ & $22.66 \pm 0.10$\\
2014-05-28 & 18:59:39 & 361.7866 & ZTSh & $R$    &  $22.71 \pm 0.08$\\
2013-07-07 & 17:18:43 & 5.6947 & AZT-22 & $B$ & $20.56 \pm 0.08$\\
2013-07-10 & 16:40:59 & 8.6581 & AZT-22 & $B$ & $20.46 \pm 0.08$\\
2013-07-12 & 17:52:31 & 10.7077 & AZT-22 & $B$ & $20.65 \pm 0.07$\\
2013-07-13 & 20:39:28 & 11.8607 & Z-1000/CrAO & $B$ & $20.51 \pm 0.25$\\
2013-07-14 & 17:29:51 & 13.6920 & AZT-22 & $B$ & $20.64 \pm 0.06$\\
2013-08-03 & 18:47:33 & 32.7642 & ZTSh   & $B$ & $22.29 \pm 0.07$\\
2014-05-28 & 20:35:56 & 361.8417 & ZTSh   & $B$ & $23.54 \pm 0.20$\\
  \hline                                       
  \end{tabular}
  \label{ourlog}
 \end{minipage}
\end{table*}
\begin{table*}
 \centering
  \caption{Reference stars from SDSS-DR9 used for the photometry. $R$ and $B$ magnitudes were obtained from $ugriz$ data using Lupton's 2005 transformation equations.}
  \begin{tabular}{ll|c|c|c|c|c|c|c}
  \hline
N & SDSS id & $u$ & $g$ & $r$ & $i$ & $z$ & $R$ & $B$ \\
\hline
1 & J142915.86+154510.0 & 17.968 (0.012) & 16.721 (0.004) & 16.228 (0.004) & 16.029 (0.004) & 15.963 (0.007) & 16.033 (0.013) & 17.095 (0.013) \\
2 & J142911.60+154535.2 & 19.808 (0.034) & 17.303 (0.005) & 15.920 (0.004) & 15.261 (0.004) & 14.933 (0.005) & 15.576 (0.013) & 17.935 (0.035) \\
3 & J142922.06+154652.2 & 20.070 (0.043) & 19.030 (0.009) & 18.715 (0.009) & 18.607 (0.011) & 18.562 (0.030) & 18.550 (0.020) & 19.356 (0.045) \\
4 & J142911.64+154807.2 & 18.994 (0.022) & 17.390 (0.005) & 16.800 (0.004) & 16.565 (0.004) & 16.433 (0.008) & 16.591 (0.013) & 17.813 (0.023) \\
5 & J142923.79+154823.4 & 21.009 (0.084) & 18.575 (0.007) & 17.627 (0.006) & 17.288 (0.006) & 17.109 (0.011) & 17.370 (0.015) & 19.132 (0.085) \\
\hline
  \end{tabular}
\label{refstars}
\end{table*}

\begin{figure}
\centering
\includegraphics[width=90mm]{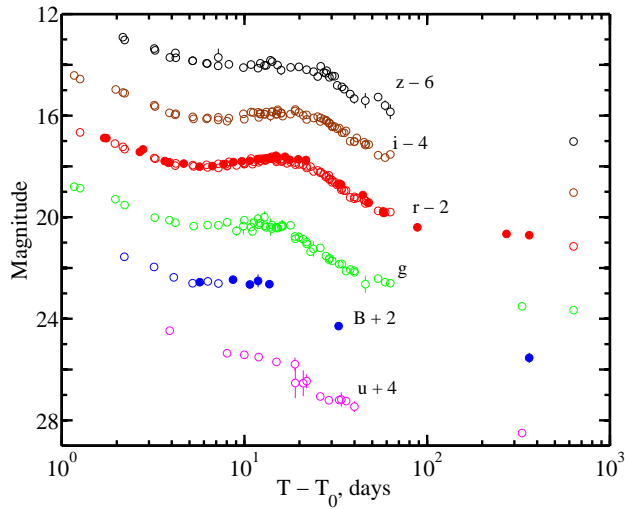}
\caption{The observed multicolour light curves of SN~2013dx connected with GRB~130702A. Filled and open circles depict the data presented in this work and collected from the literature, respectively (see Section~\ref{sec:otherdata}). The supernova phenomenon is clearly seen in every filter. The afterglow component and the modelled supernova contribution in $r$ filter are shown by the dashed and dotted lines, respectively (see Section~\ref{sec:lc} for details). The Galactic extinction is not taken into account.}
\label{lcobs}
\end{figure}

We started to monitor the field of GRB~130702A with 1.5-m AZT-22 telescope 
of Maidanak astronomical observatory on July 3, 2013, i.e. one and a half days after the
\textit{Fermi} trigger \citep{gcn14988}. We took 6 frames in an $R$-filter with exposures of 
300~s each. The afterglow reported by \citet{singer-gcn} was clearly detected
on a stacked frame. We continued the observations of the afterglow and rising
supernova taking several 300-s frames in $R$-band on July 4-12, 14-16, 18, 
19, 21, 23 and August 15, 18, 28. We also took observations in $B$-band taking
frames with the same exposures on July 7, 10, 12, 15.

The OA was observed by 0.7-m AS-32 telescope of Abastumani Astrophysical Observatory 
taking unfiltered images with exposures of 120 s on July 3, 4, 15 and 16. The 1-m Zeiss-1000 (Z-1000)
telescope of Simeiz branch of the Crimean Astrophysical Observatory took several 120-s frames
of the OA and emerging supernova in $B$ and $R$ filters on July 13. The observations with this 
instrument were continued on July 15-16 only in $R$ filter with the same exposures.

The 2.6-m Shajn telescope (ZTSh) of the Crimean Astrophysical Observatory observed of the OA
taking 1-h set of 60-s images in $R$ filter on July 5. The instrument also observed the decaying 
supernova with 120-s frames taken on August 3 ($B$ and $R$) and August 4 (only $R$).

The late phase of supernova decay was observed by 1.5-m Russian-Turkish telescope (RTT-150) at Tubitak
observatory on August 28 (3 frames of 300 s in $R$-band) and by the 2.5-m Nordic Optical Telescope (NOT) at
Roque de los Muchachos Observatory on September 27-28.

We observed the host galaxy on May 28, 2014
with ZTSh taking 81 images in $R$ filter and 54 images in $B$ filter
with exposures of 60 s. However, the seeing in observations in R filter was not good enough
to clearly separate the host galaxy flux from that of the neighbour
big galaxy.

We also obtained images of the host galaxy on March 30, 2014 with the 1.5-m AZT-33IK telescope of the Sayan observatory (Mondy)
in  $R$ filter with total exposure of 2 h. 
The log of all observations is presented in Table~\ref{ourlog}.

All our optical data were processed using NOAO's \textsc{iraf} software package\footnote{\textsc{iraf} is the Image Reduction and Analysis Facility, a general purpose software system for the reduction and analysis of astronomical data. \textsc{iraf} is written and supported by the National Optical Astronomy Observatories (NOAO) in Tucson, Arizona. NOAO is operated by the Association of Universities for Research in Astronomy (AURA), Inc. under cooperative agreement with the National Science Foundation. \url{http://iraf.noao.edu/ } }. Standard image processing (bias, dark reduction, flatfielding) was done using the task \texttt{ccdproc}. Some images were combined in sums using the tasks \texttt{imlintran} and \texttt{imcombine} to provide better signal-to-noise ratio. The aperture photometry was done using the \textsc{apphot} subpackage, using an aperture radius of twice the FWHM that measured for point sources on each night. The photometry is based on reference stars from SDSS-DR9 listed in Table \ref{refstars} ($R$ and $B$ magnitudes are transformed from $ugriz$ system using Lupton (2005) transformation equations\footnote{\url{http://classic.sdss.org/dr6/algorithms/sdssUBVRITransform.html\#Lupton2005 }}). The unfiltered data from Abastumani AS-32 telescope were calibrated using the same $R$ magnitudes and the additional correction constant of $+0.034$ magnitudes determined by the spectral properties of the AS-32 CCD and Johnson $R$ filter and calculated using numerous optical observations of other GRB optical afterglows.

All the data reported in Table~\ref{ourlog} supersede the data previously published in GCN circulars (NN 14988, 14996, 15003, 15243).

\subsubsection{Data from other sources}
\label{sec:otherdata}

We collected all available data on GRB~130702A and SN 2013dx observations, published up to now. 
We took 87 photometrical points from D15 in $uUgri$  filters and 192 points from T16
in $g'r'i'z'$ filters. We added 6 points from \citet{singer_paper} in $B$ filter but due to the lack
of numerical data in explicit form we estimated the corresponding values basing on the figure 3 of~\citet{singer_paper} with the magnitude uncertainties $0\fm1$. 

We also adopted the host galaxy magnitudes obtained by T16 in $u'g'r'i'z'$ filters at $t = 632$~d 
and in $R$ filter at $t = 330$~d after the trigger. 

We converted $U$ values from D15 and our $R$ magnitudes from the Vega to the AB system and then to $u$ and $r$, respectively, using methods from \citet{vegaAB}. We also converted all magnitudes used for the light curves construction from $u'g'r'i'z'$ to $ugriz$ system using SDSS photometric equations\footnote{\url{http://classic.sdss.org/dr7/algorithms/jeg_photometric_eq_dr1.html}}. The observed multicolour light curves of the GRB~130702A optical counterpart are presented in Fig.~\ref{lcobs}.

\subsection{Radio observations}

The afterglow of GRB~130702A was observed at mm-wavelengths with the Combined Array for Research in Millimeter Astronomy (CARMA) on 2013 July 4, i.e. 2 d after the trigger, at a frequency of 93 GHz (a wavelength of 3 mm) \citep{mm}. A source coincident with the optical counterpart was discovered with a flux density $\sim2$ mJy.

The afterglow was also observed by the Westerbork Synthesis Radio Telescope at 
4.9 GHz 2.56-3.05 d after the burst \citep{radio}. A radio source with a flux 
density of $1.23 \pm 0.04$ mJy was detected at the position of the optical counterpart.
The radio counterpart was also observed with the Karl G. Jansky Very
Large Array in $C$-band 2.29 d after the GBM trigger \citep{radio2}. The radio source
was detected with a flux density of 1.49 mJy at 5.1 GHz, and 1.60 mJy at 7.1 GHz.

We observed the position of the radio afterglow using the 22-m radio-telescope RT-22 of the 
Crimean Astrophysical observatory at 36 GHz on July 5 and 6, 2013, i.e. 3.69 and 4.68 d
after the trigger. At the position of the afterglow we obtained 2-$\sigma$ upper limits of $0.6$~mJy
and $0.5$~mJy respectively.

A summary of the GRB~130702A afterglow properties is collected in Table~\ref{grb}.

\begin{table}
 \centering
  \caption{Summary of the GRB~130702A and its afterglow properties.}
  \begin{tabular}{ll}
  \hline
Discovered by & GBM/\textit{Fermi} \\
$T_0$ & 00:05:23.079 \textsc{ut} on July 2, 2013 \\
$\alpha$, $\delta$ (GBM) & $218\fdg81$, $+12\fdg25$, $4\degr$ radius \citep{collazzi}\\
$\alpha$, $\delta$ (LAT) & $216\fdg4$, $+15\fdg8$, $0\fdg5$ radius \citep{cheung}\\
$T_{90}$ (GBM) & 59 s \citep{collazzi}\\
$T_{90}$ (SPI-ACS) & $545 \pm 60$ s \\
Highest energy & 1.5 GeV \citep{cheung}\\
z & 0.145 \citep{z1,z2,z3}\\
$t_{\rm jb}^o$ & $1.17 \pm 0.09$ d \citep{singer_paper}\\
$t_{\rm jb}^X$ & $1.7 \pm 0.2$ d \\
$\alpha_2$ & $1.27 \pm 0.02$ \\
$E_{\rm iso, \gamma}$ & $(6.6 \pm 0.4) \times 10^{50}$ erg \\
$E_{\rm iso, \it X}$ & $> 2.7 \times 10^{50}$ erg \\
  \hline
\end{tabular}
\newline $^o$ - obtained using optical data;
\newline $^X$ - obtained using X-ray data;
\label{grb}
\end{table}

\subsection{Host galaxy}

\begin{figure}
\centering
\includegraphics[width=90mm]{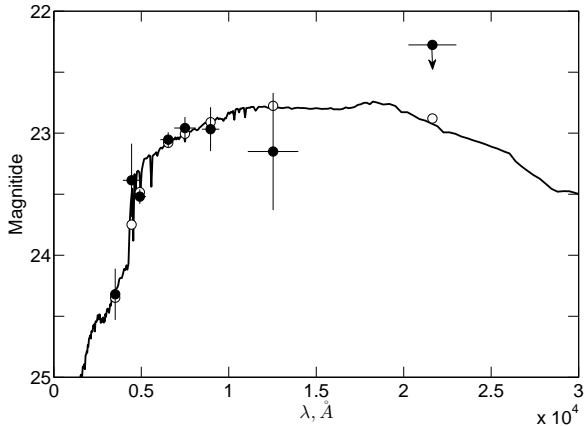}
\caption{The SED (line) of the host galaxy of GRB~130702A fitted by the \textit{Le Phare} with fixed redshift. Filled circles depict the data points in $u'g'r'i'z'JK_s$ filters, respectively, taken from T16 and from ZTSh observations. Open circles show the model magnitudes for each filter.}
\label{sed}
\end{figure}

\citet{kelly} reported that GRB~130702A occurred in a dwarf satellite of a massive galaxy at a redshift $z=0.145$ and provided some physical properties of the galaxy using a SED based on the SDSS $ugriz$ photometry. As input parameters we took the $u'g'r'i'z'JK_s$ magnitudes of the host galaxy from T16, $B$ magnitude from ZTSh (see Table~\ref{ourlog}), and fixed redshift and fitted the SED of the galaxy using the \textsc{Le~Phare} software package~\citep{lephar1,lephar2}. We used the \textsc{PEGASE2} population synthesis models library \citep{pegase} to obtain the best-fitted SED and the main physical parameters of the galaxy: age, mass, and star formation rate. The apparent magnitudes used for the fitting are listed in Table~\ref{host}.

The host galaxy of GRB~130702A ($\chi^2/$DOF$ = 2.9/8$) is fitted by the SED of an irregular dwarf galaxy with the age of $4.3 \times 10^{9}$ yr and the mass of $M_{\rm host} = 1.3 \times 10^{8}~\rm M_{\sun}$, which is slightly higher than the mass reported by \citet{kelly}. We also obtained the negligible mean extinction in the host $A_V^{\rm host} = 0$ with the best extinction law of the SMC \citep{smc_prevot} and the star formation rate of $SFR_{\rm host} = 0.05~\rm M_{\sun}/$yr. The two latter results are in a good agreement with those of the previous studies \citep{kelly}. The best fitted SED of the host galaxy is shown in Fig.~\ref{sed}. A summary of the host galaxy properties are collected in Table~\ref{host}.

\begin{table}
 \centering
  \caption{A summary of the GRB~130702A host galaxy properties, the magnitudes are in AB system corrected for the expected Galactic foreground extinction.}
  \begin{tabular}{ll}
  \hline
type & irregular dwarf \\
$M_r$ & $-16.2^m$\\
$u'_{\rm host}$ & $24.32 \pm 0.21$ (T16)\\ 
$B_{\rm host}$ & $23.39 \pm 0.20$\\ 
$g'_{\rm host}$ & $23.52 \pm 0.06$ (T16)\\
$r'_{\rm host}$ & $23.05 \pm 0.06$ (T16)\\
$R_{\rm host}$ & $23.17 \pm 0.06$ (T16)\\
$i'_{\rm host}$ & $22.96 \pm 0.09$ (T16)\\
$z'_{\rm host}$ & $22.97 \pm 0.18$ (T16)\\
$J_{\rm host}$ & $23.15 \pm 0.48$ (T16)\\
$K_{s, \rm host}$ & $> 22.28$ (T16)\\
Age & $4.3 \times 10^{9}$ yr\\
$M_{\rm host}$ & $1.3 \times 10^{8}~\rm M_{\sun}$\\
$A_V^{\rm host}$ & $\sim 0$\\ 
$SFR_{\rm host}$ & $0.05~\rm M_{\sun}/$yr\\
  \hline
  \end{tabular}
\label{host}
\end{table}

\section{SN light curve}
\label{sec:lc}
%
\begin{figure*}
\centering
\begin{minipage}{1\linewidth}
\begin{tabular}{cc}
\includegraphics[width=0.5\textwidth]{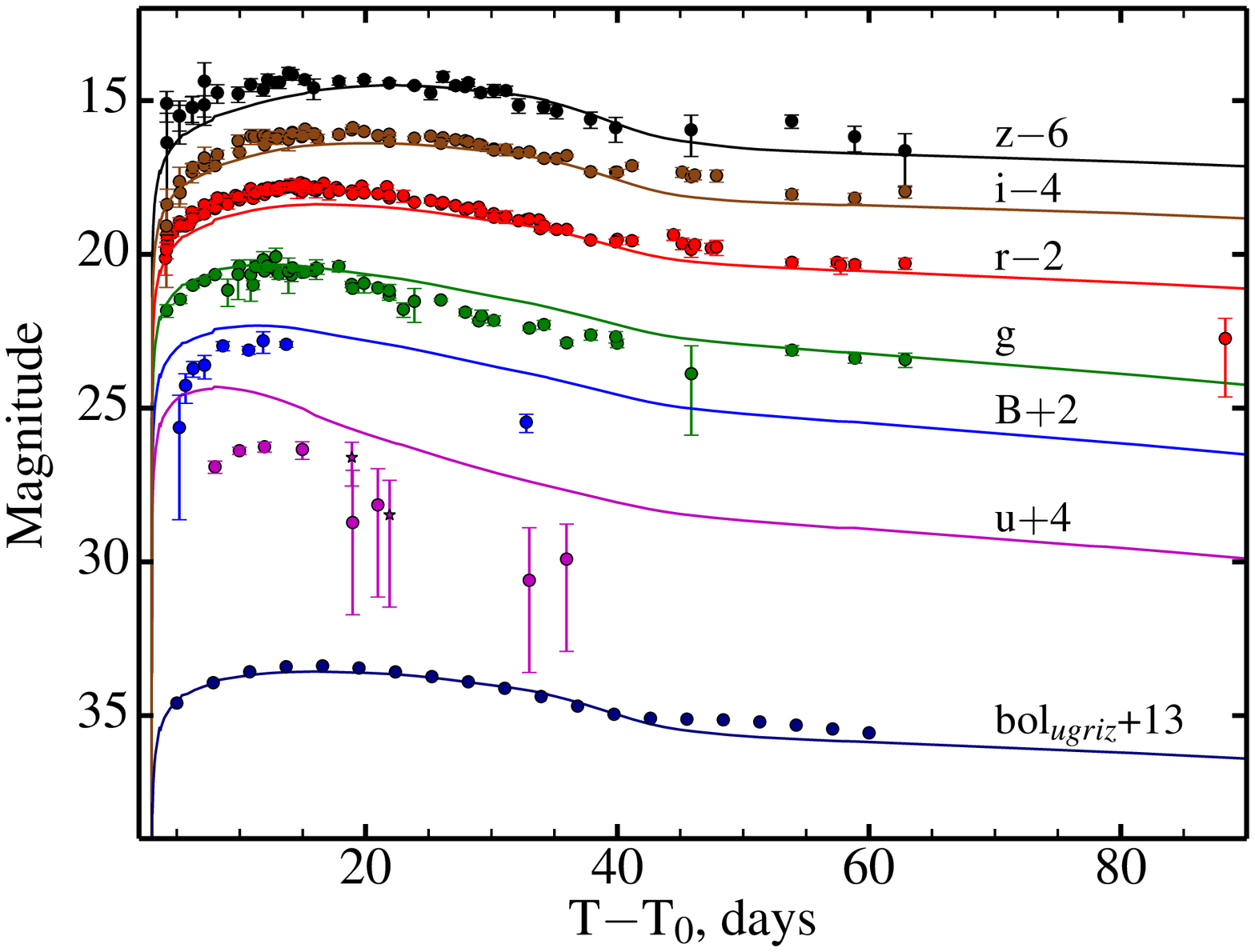} & \includegraphics[width=0.5\textwidth]{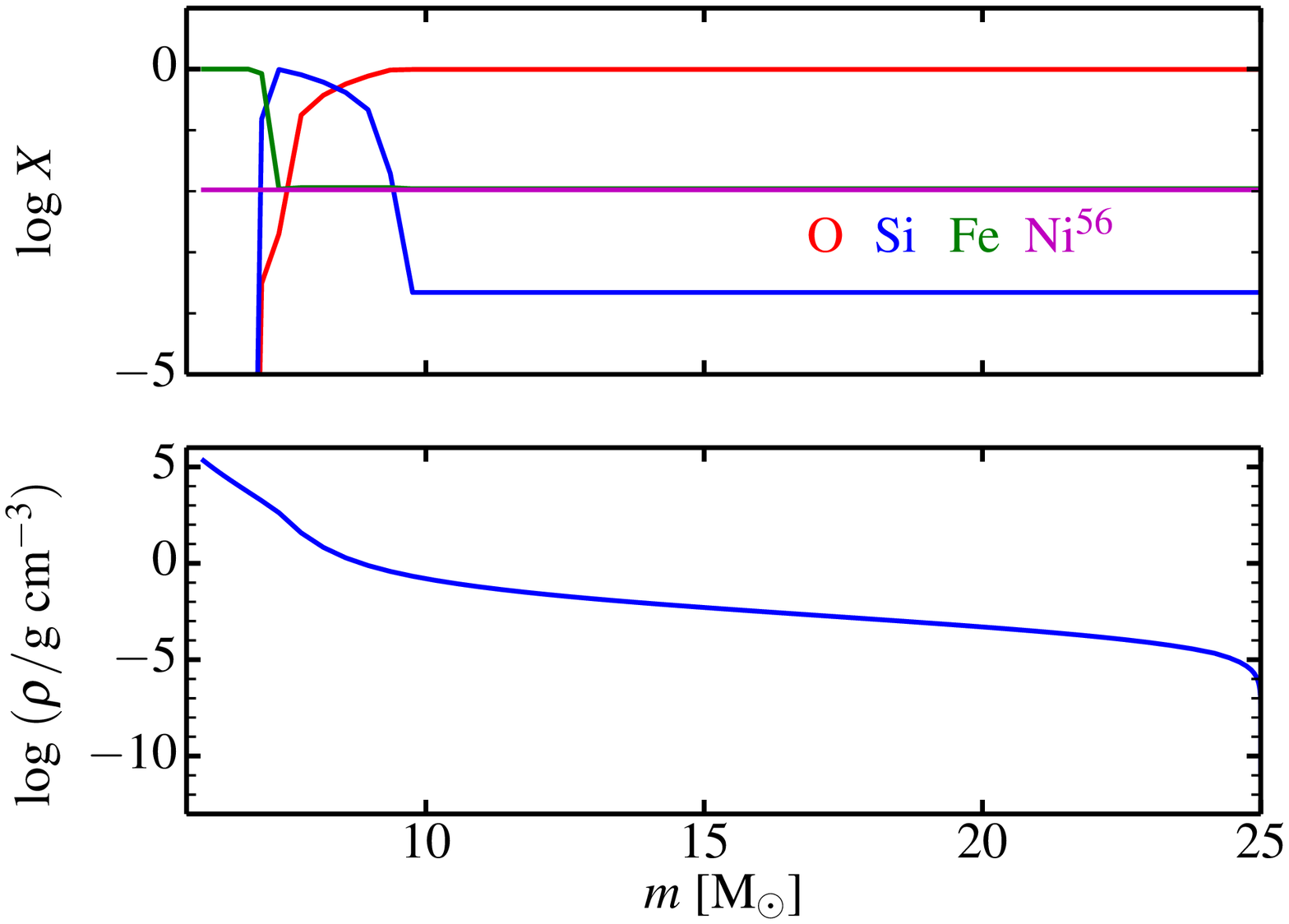} \\
a) & b) \\
\includegraphics[width=0.5\textwidth]{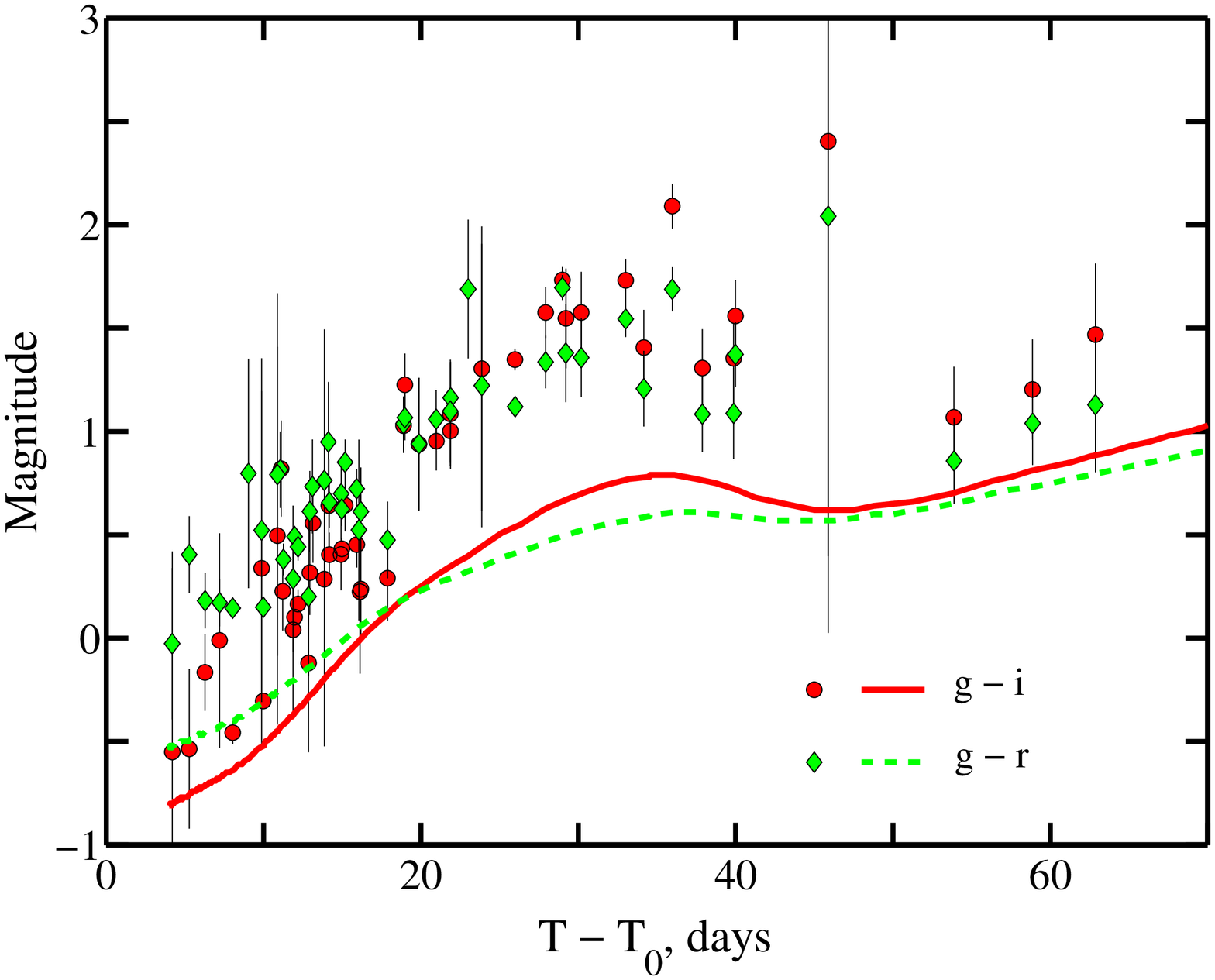} & \includegraphics[width=0.5\textwidth]{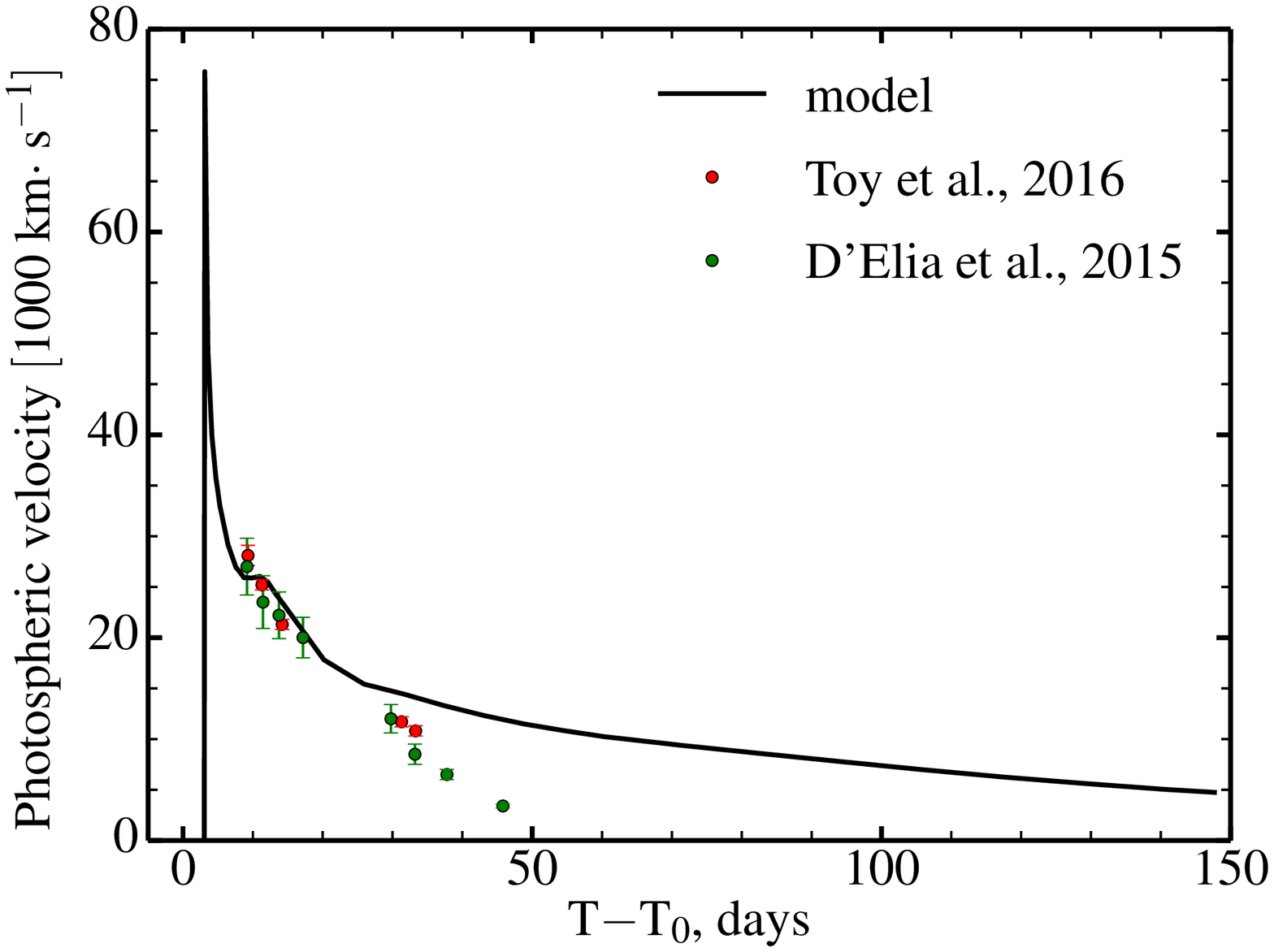} \\
c) & d) \\
\end{tabular}
\caption{a) The multicolour light curves of SN~2013dx. The Galactic extinction, the flux 
from the host galaxy and the optical afterglow contribution are excluded. 
Solid lines show the best model of the SN light curve obtained by \textsc{stella}
(see Section~\ref{sec:modelling}). The quasibolometric light curve of the SN in AB photometric system obtained as a sum of the fluxes in $ugriz$ filters, is marked as $\textrm{bol}_{ugriz}$. The data and model are in observer frame. b) The mass fraction of O, Si, Fe, and $\rm^{56}$Ni in the ejecta and density profile for the optimal pre-supernova star model with respect to interior mass.  The central region of 6~$\rm{M_{\sun}}$ is taken away. c) Colour evolution of SN~2013dx. Points and lines show the variation of colour indexes $g-i$ 
(red circles, red solid line) and $g-r$ (green diamonds, green dashed line) with time for observations and optimal model, consequently. d) The evolution of photospheric velocities of SN~2013dx measured 
via observations (points) and calculated from the modelling (solid line). 
The plot is in observer frame.}
\label{lightcurve}
\end{minipage}
\end{figure*}

To construct a light curve of a SN associated with a GRB one needs to take into account 
the contribution from GRB OA, GRB host galaxy, and the light extinction in
the Galaxy and in the host galaxy. 

First, all photometrical data of GRB~130702A were corrected for Galactic extinction using the extinction maps from
\citet{schlafly} with $E(B-V) = 0.038$.

Second, the contribution of the host galaxy was eliminated by flux subtraction. We used data from T16 for $ugriz$ filters (see Section~\ref{sec:otherdata}). For the $B$-filter we used the value calculated via Lupton (2005) transformation equations from $ugr$ values of T16.

To distinguish the optical afterglow from the SN we considered the approximation by the broken power-law slope obtained from X-ray afterglow light curve (see Section~\ref{sec:xrt}) and assumed that the decay of the afterglow is achromatic from X-ray to optical domain after the jet-break (see Section 4.1 in T16), and thus we fitted the optical data obtained between $t_b$ and $t = 4$ d after the trigger with a single power law with fixed $\alpha_X = -1.27$ (see Section~\ref{sec:xrt}) and then subtracted calculated flux from all photometrical points.

T16 estimated the host extinction as $A_V = 0.13 \pm 0.23$ by fitting the SED of the galaxy 
with a simple power-law model. This value is consistent with zero within the 1-$\sigma$ confidence interval, 
moreover, our modelling of the host galaxy SED showed that the mean host extinction is negligible, 
so we assumed $A_V^{\rm host} = 0$ and did not to take it into account for further modelling. 

Finally, we converted all values to AB magnitudes and constructed the
optical light curves of SN~2013dx, which are shown in Fig.~\ref{lightcurve}a. We also constructed
the quasibolometric light curve of the SN in the AB photometric system by taking the sum of the fluxes in $ugriz$ filters. We used the code by \citet{ishida_desouza,ishida2}\footnote{\url{https://github.com/emilleishida/snclass}} that adapts Gaussian Processes for interpolating the light curves. Due to the fact that the data coverage in the $u$ filter is less than the others, we had to interpolate the $u$ light curve between 20 and 35 d and to extrapolate after $t=35$ d. We assumed that the $u-g$ colour corresponding to the last day of observations in $u$ filter remains unchanged up to $t = 60$ d. The quasibolometric light curve is also shown in Fig.\ref{lightcurve}a.
%
\section{Modelling of the SN light curve}
\label{sec:modelling}

\subsection{Modelling}

\begin{figure*}
\centering
\begin{minipage}{1\linewidth}
\begin{tabular}{cc}
\includegraphics[width=0.5\textwidth]{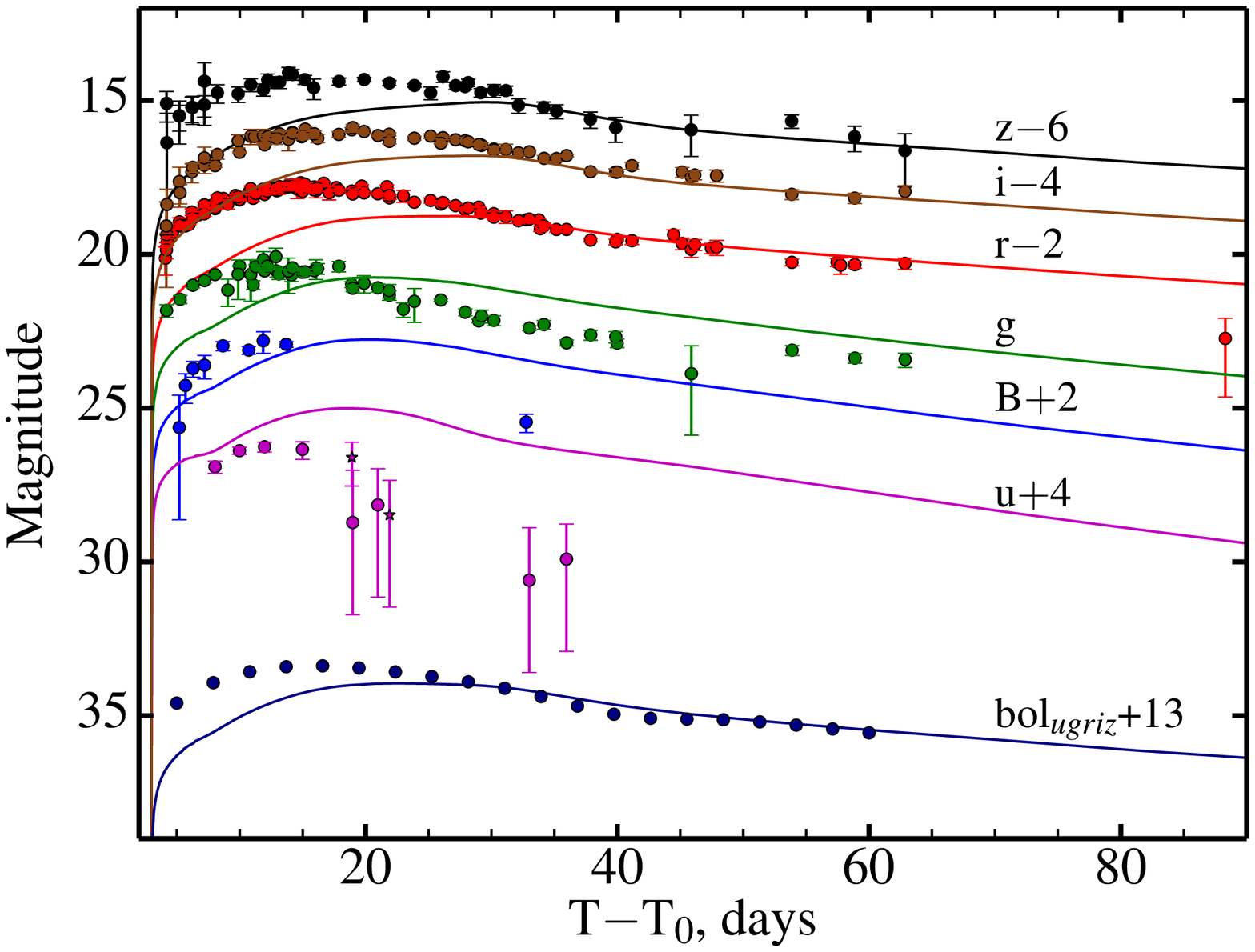} & \includegraphics[width=0.5\textwidth]{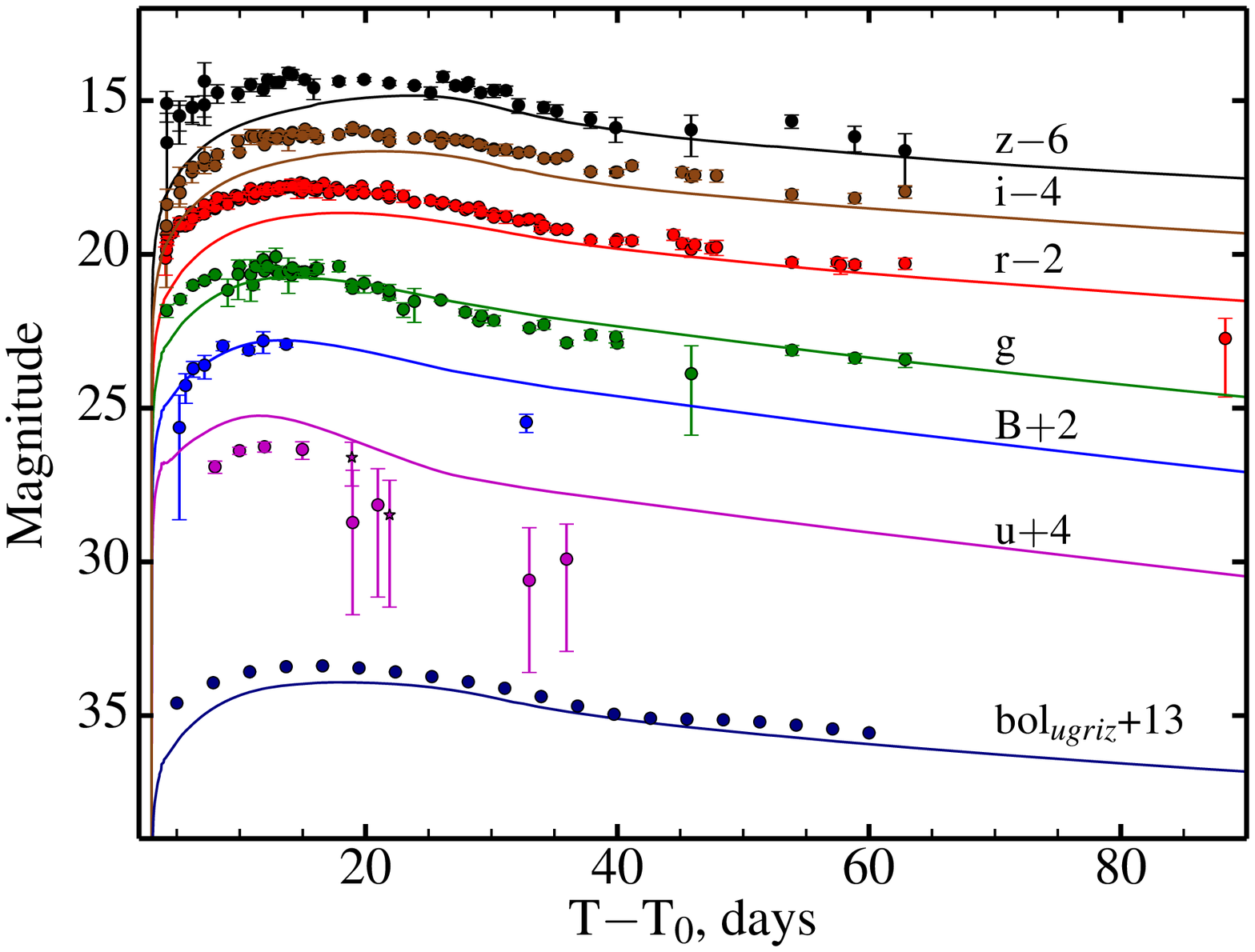} \\
a) & b) \\
\end{tabular}
\caption{The result of modelling (lines) the observed light curves of the SN~2013dx (points) with $M_{\rm ej}=3.1$~$\rm{M_{\sun}}$, $M_{\rm^{56}Ni} = 0.37$~$\rm{M_{\sun}}$, and $E_{K} = 8.2\times10^{51}$ erg corresponding to T16. Two different distributions of $\rm^{56}$Ni in the ejecta are considered: a) totally mixed, b) not mixed. The quasibolometric light curve of the SN in AB photometric system obtained as a sum
of the fluxes in $ugriz$ filters, is marked as $\textrm{bol}_{ugriz}$.}
\label{lc-toy}
\end{minipage}
\end{figure*}

The numerical multicolour light-curve calculations for SN~2013dx were performed with 
the one-dimensional multi-frequency radiation hydrodynamics code 
\textsc{stella}~\citep{Blinnikov98,Blinnikov06}. In the current calculations we 
adopted 100 zones for the Lagrangean coordinate and 130 frequency bins. 
As input parameters we varied the pre-supernova star mass $M$ and radius $R$, the energy of the 
outburst $E_{\rm oburst}$, the mass of synthesized nickel $M_{\rm^{56}Ni}$, the mass of the 
resulting compact remnant $M_{\rm CR}$, and the the initial distribution of chemical 
elements in the pre-supernova star.

The mass $M_{\rm CR}$ in the central part of a pre-supernova star with a fixed
radius (which is much less than the outer radius of the star) is
treated as a point-like source of gravity which has a non-negligible influence on
expansion of the innermost layers of SN-ejecta.
The explosion is initiated by putting thermal energy to the 
innermost layers. The ejecta of SNe has the same chemical composition as pre-supernova star except for 
$\rm^{56}$Ni, because we do not follow the explosive nucleosynthesis. $\rm^{56}$Ni can 
be put in the centre of SN~ejecta in the calculations as well as be spread out within any region.

The mass and radius of pre-supernova star for the model, which offers the best agreement 
with the observations of SN~2013dx, is $M = 25$ $\rm{M_{\sun}}$ and 
$R = 100$ $\rm{R_{\sun}}$. The resulting energy of explosion is 
$E_{\rm oburst} = 3.5 \times 10^{52}$ erg. A part of the outburst energy that went into the radiation from far IR to extreme UV (in the range from $6.3 \times 10^{13}$ Hz, or 4.8 $\mu$m, to $1.7 \times 10^{16}$ Hz, or 70 eV), i.e. bolometric energy $E_{\rm bol}$ in time interval of 126 days after the trigger was $3.1 \times 10^{49}$ erg. The $0.2$ $\rm{M_{\sun}}$ of $\rm^{56}$Ni is 
totally mixed through the ejecta. The central region becomes a black hole with 
$M_{\rm CR} = 6 \rm{M_{\sun}}$. The parameters $E_{\rm oburst}$ and $M_{\rm^{56}Ni}$ were adopted from D15, 
the rest of the parameters were picked up to better match the observed light curve by 
the model. 

Fig.~\ref{lightcurve}a compares the optical ($ugriz$ and $B$) light curves of the 
model with the observations of SN~2013dx. From the observational data in $ugriz$ 
filters, the quasibolometric light curve was also derived and compared with the modelled 
one. The quasibolometric light curve of the SN in AB photometric system is obtained as a sum
of the fluxes in $ugriz$ filters. To match better the observational and modelled light curves, the global time offset (3 days), i.e. a simultaneous time-shift of modelled light curves relative to $T_0$, has been applied. 
Fig.~\ref{lightcurve}a shows that the modelled light curves match adequately the 
observations from $z$ to $g$ filters, especially for quasibolometric one. 
However, in the blue filters ($B$ and $u$) the model and observations disagree. 
We discuss a possible nature of this behaviour in Section~\ref{sec:discussion}.
Also we obtained the value of $t_{\rm peak}$, which is a time of maximum for quasibolometric modelled light curve relative to the trigger time $T_0$.

The distribution of chemical elements and density profile for a pre-supernova star are shown 
in Fig.~\ref{lightcurve}b. It is worth noting, that the model shows a total absence 
of hydrogen and helium in the pre-supernova star composition which is common for Type Ic 
supernovae~\citep{Filippenko1997}. 

In Fig.~\ref{lightcurve}c we plot the colour evolution of the SN, and the 
Fig.~\ref{lightcurve}d shows observer-frame photosperic velocities in comparison 
with observational data obtained via spectroscopy by D15 and T16. The model 
evolution of the photospheric velocities is in good agreement with direct 
observations during the first $\sim$30 days after the explosion. The summary of the 
SN~2013dx parameters is collected in Table~\ref{sn}. It also shows the comparison between the parameters of the SN obtained by other studies.

Following the description of supernova explosion parameters in T16, we performed the LC modelling adopting $M_{\rm^{56}Ni} = 0.37$~$\rm{M_{\sun}}$ (located in the centre of explosion and no mixing), $E_{\rm K} = 8.2\times10^{51}$ erg, and $M_{\rm ej}=3.1$~$\rm{M_{\sun}}$. In the current model we assumed, following T16, that there is no compact object at all, i.e. no gravitational centre that affects the expansion of the innermost layers. Since in T16 it is written that the initial radius before explosion is small, we put $R=10~\rm R_{\sun}$. 
Fig.~\ref{lc-toy}a represents the comparison between SN~2013dx observations and \textsc{stella} model for T16 SN parameters. In the early phases the modelled light curves lie below the observational ones in all filters. Starting from $\sim$30 day after the explosion $z$, $i$, $r$, and quasibolometric modelled light curves fit well the observational data. We also made a model keeping absolutely the same parameters as in T16 but with nickel mixed totally in the ejecta as in our optimal model. The nickel mixing partly improved the increasing part of modelled light curves making them a bit more consistent with observations (see~Fig.~~\ref{lc-toy}b). The global time offset (3 days) has been applied to all modelled light curves. 

\begin{table*}
\centering
 \begin{minipage}{140mm}
  \caption{A summary of the SN~2013dx parameters obtained by \textsc{stella} in comparison with those of D15 and T16. All masses are in $\rm{M_{\sun}}$.}
  \begin{tabular}{l|l|l|l}
  \hline
Parameter & \textsc{stella} model & D15 & T16 \\
  \hline
$M$ & 25 & $ \sim 25-30^a$ & -- \\ 
$M_{\rm CR}$ & 6 & -- & -- \\
$E_{\rm oburst}$ & $3.5 \times 10^{52}$ erg & $\sim 3.5 \times 10^{52}$ erg & $(8.2 \pm 0.4) \times 10^{51}$ erg\\
$M_{\rm ej}$ & $19$ & $\sim 7$ & $3.1 \pm 0.1$ \\
$R$ & $100~\rm R_{\sun}$ & -- & -- \\
$M_{\rm^{56}Ni}$ & 0.2, totally mixed & $\sim 0.2$ & $0.37 \pm 0.01$ \\
$M_{\rm O}$ & 16.6 & -- & -- \\
$M_{\rm Si}$ & 1.2 & -- & -- \\
$M_{\rm Fe}$ & 1.2 & -- & -- \\
$E_{\rm bol}$& $3.1 \times 10^{49}$ erg & -- & -- \\
$t_{\rm peak}$ & $14.35^{\ast}$ d & $15 \pm 1^{\ast\ast}$ d & $13.2 \pm 0.3^{\ast\ast}$ d \\
  \hline
  \end{tabular}
\newline $^a$ - mass of the progenitor on the main sequence
\newline $^{\ast}$ - on the bolometric light curve
\newline $^{\ast\ast}$ - on the light curve in the filter $r$
\end{minipage}
\label{sn}
\end{table*}

\subsection{Influence of different initial parameters}

\begin{figure*}
\centering
\begin{minipage}{1\linewidth}
\begin{tabular}{cc}
\includegraphics[width=0.5\textwidth]{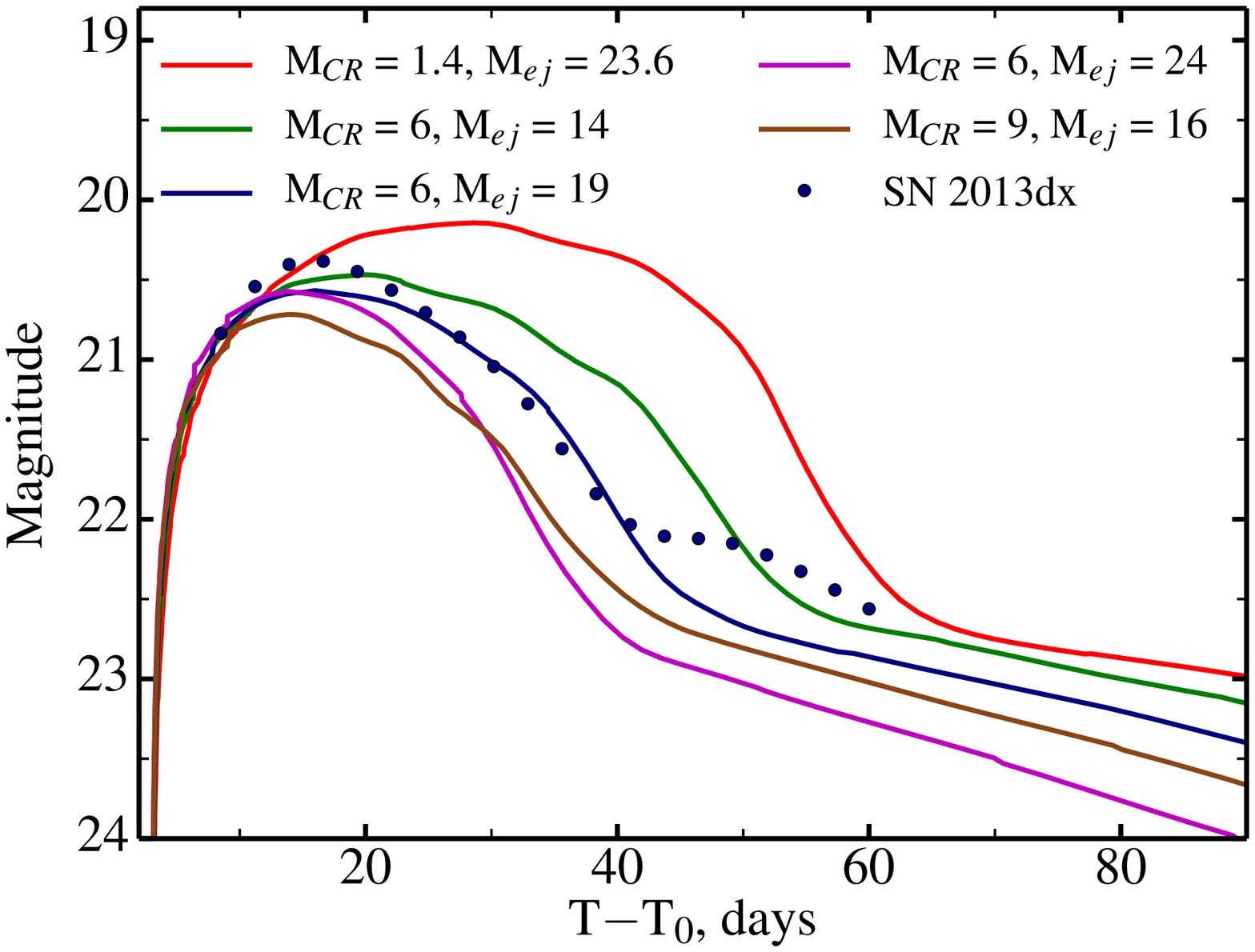} & \includegraphics[width=0.5\textwidth]{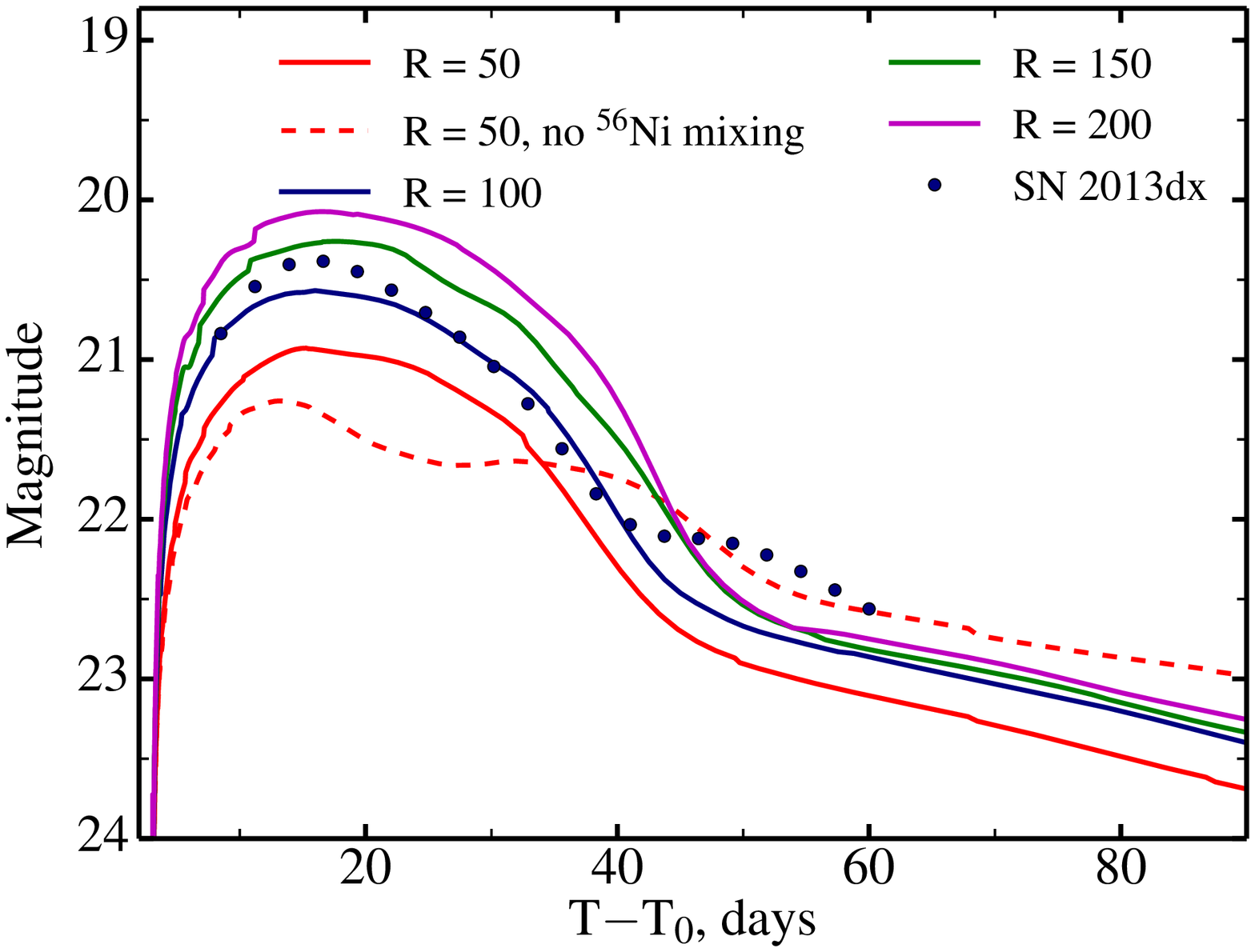} \\
a) & b) \\
\includegraphics[width=0.5\textwidth]{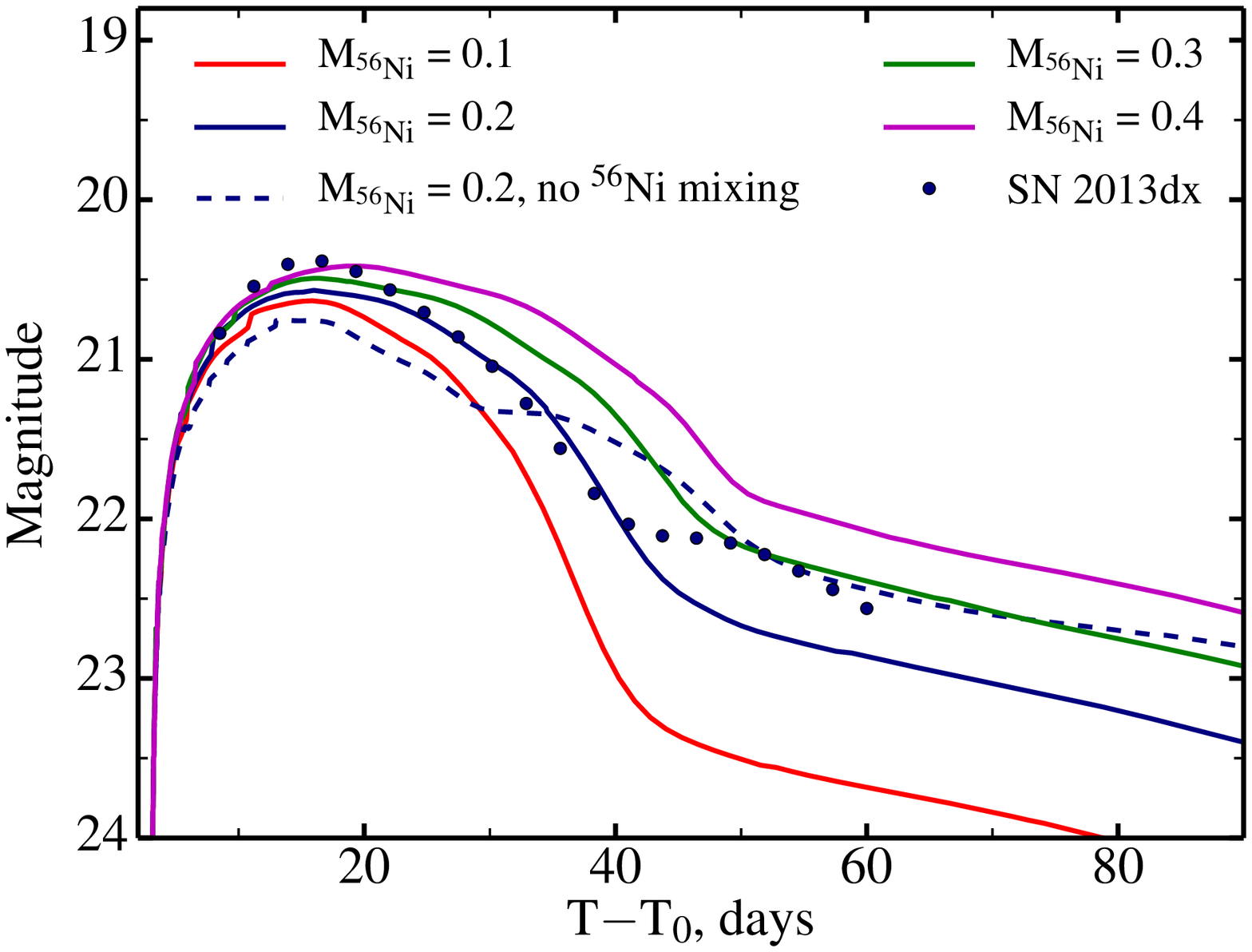} & \includegraphics[width=0.5\textwidth]{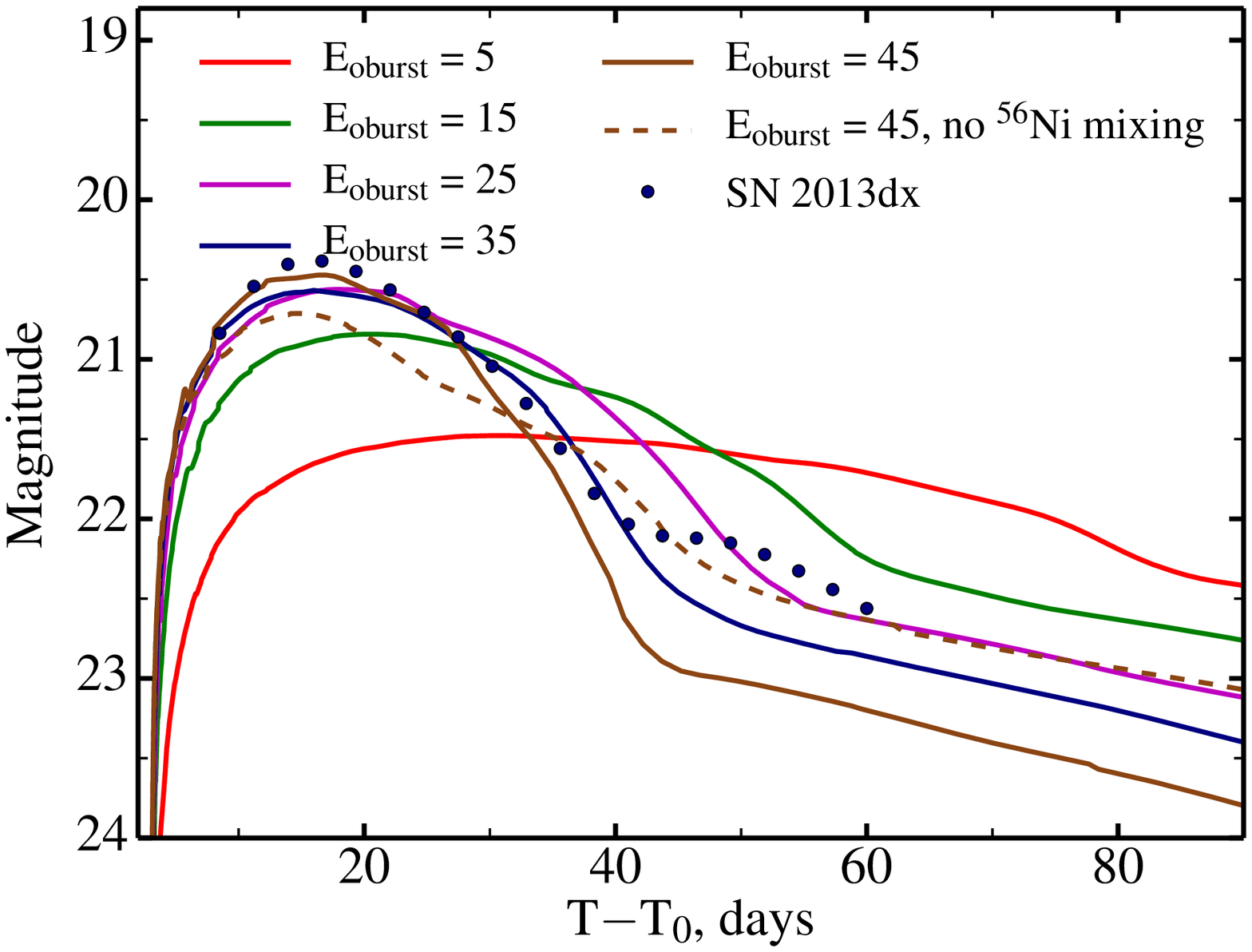} \\
c) & d) \\
\end{tabular}
\caption{The dependence of the quasibolometric light curve ($ugriz$ filters) on the different 
parameters of the optimal model (dark blue curve): a) the mass of the 
pre-supernova star and its distribution between the ejecta and the compact remnant (here and in other panels masses are in units of Solar mass); 
b) the radius of the pre-supernova star in units of Solar radius; c) the mass of the synthesized~$\rm^{56}$Ni; d) the energy of the outburst in 
units of $10^{51}$ erg. Filled circles 
show the observational quasibolometric light curve of the SN~2013dx. $\rm^{56}$Ni is totally mixed through the ejecta for all presented models except the models shown with dashed line.}
\label{sens}
\end{minipage}
\end{figure*}

In this section we consider the dependence of quasibolometric light curve from some input physical parameter of the model while the others remain fixed. We varied the mass $M$ and the radius $R$ of the pre-supernova star, the mass of synthesized $\rm^{56}$Ni, and the energy of the burst $E_{\rm oburst}$. All plots in comparison with observational quasibolometric light curve are presented in Fig.~\ref{sens}.

Fig.~\ref{sens}a,b,c demonstrate that the ejecta mass affects the descending part of light curve, dependence of radius is stronger on the domes of light curve, however, the tail is mainly determined by $\rm^{56}$Ni abundance. The models are brighter for higher ejecta or $^{56}$Ni mass, and larger radius. The decrease of the compact remnant mass provides wider maximum of the light curve. In Fig.~\ref{sens}b,c,d we also present the models with the same amount of $\rm^{56}$Ni as our optimal model (0.2~$\rm M_\odot$) but with and without mixing through the ejecta. When $\rm^{56}$Ni is mixed, the light-curve maximum is brighter. Moreover, the gamma photons from radioactive decay are not trapped inside an envelope and do not heat the ejecta. Therefore, a photosphere goes faster to the centre and the SNe becomes dimmer on the tail. 

Decreasing the outburst energy makes the diffusion time of gamma-photons from radioactive decay longer.
This manifests itself as a widening of the light curve, moreover, the diminution of decline rate occurs later (see Fig.~\ref{sens}d).
Approximate limits of the main model parameters can be derived based on the Figure~\ref{sens}. The pre-supernova star mass and radius varies in the range 23--27 $\rm M_\odot$ and 75--125 $\rm R_\odot$, respectively; the $\rm^{56}$Ni mass lies between 0.15 and 0.25 $\rm M_\odot$; and the explosion energy $E_{\rm oburst}$ --- in the range of (30--40) $\times 10^{51}$ erg.

\begin{figure*}
\centering
\begin{minipage}{1\linewidth}
\begin{tabular}{cc}
\includegraphics[width=0.5\textwidth]{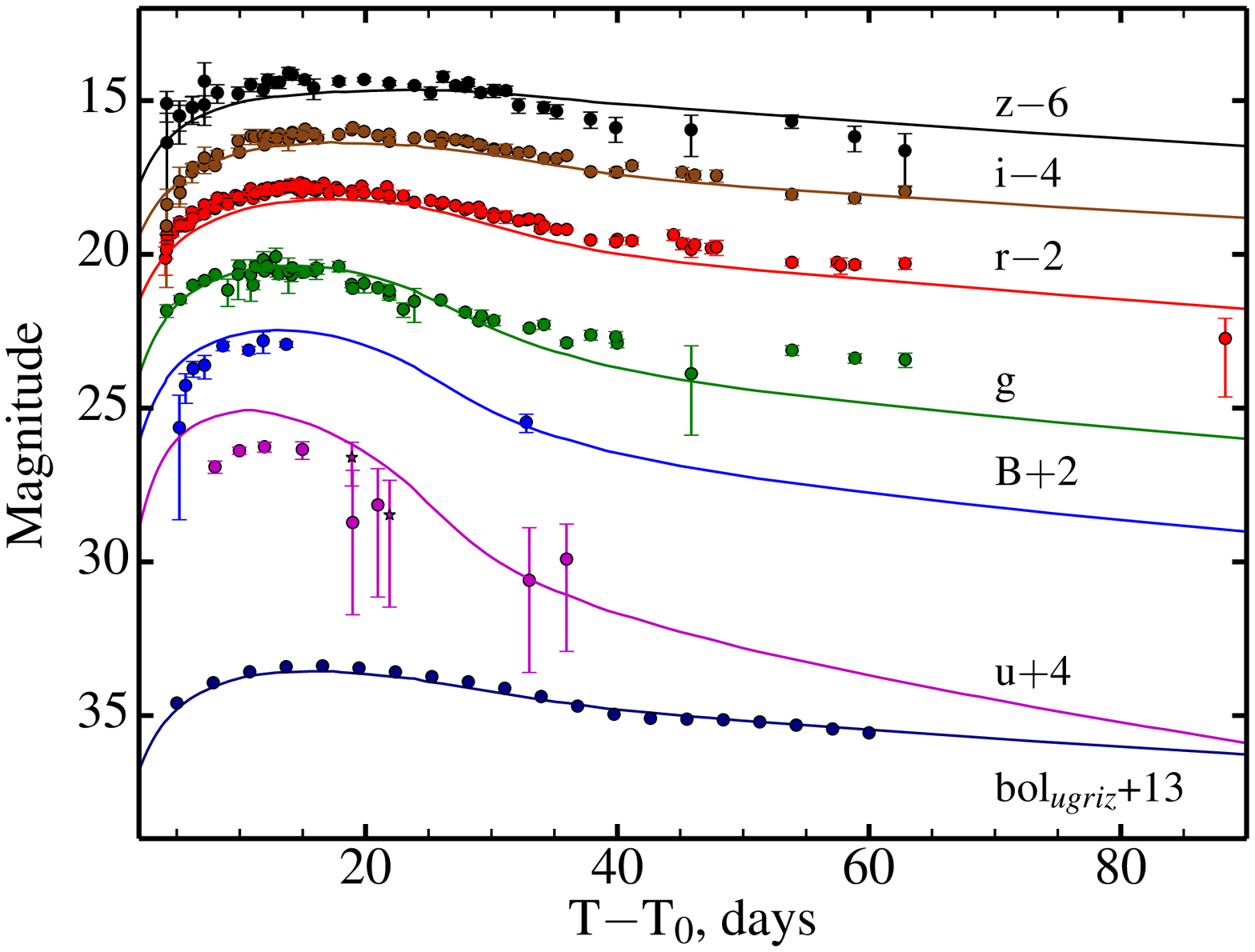} & \includegraphics[width=0.5\textwidth]{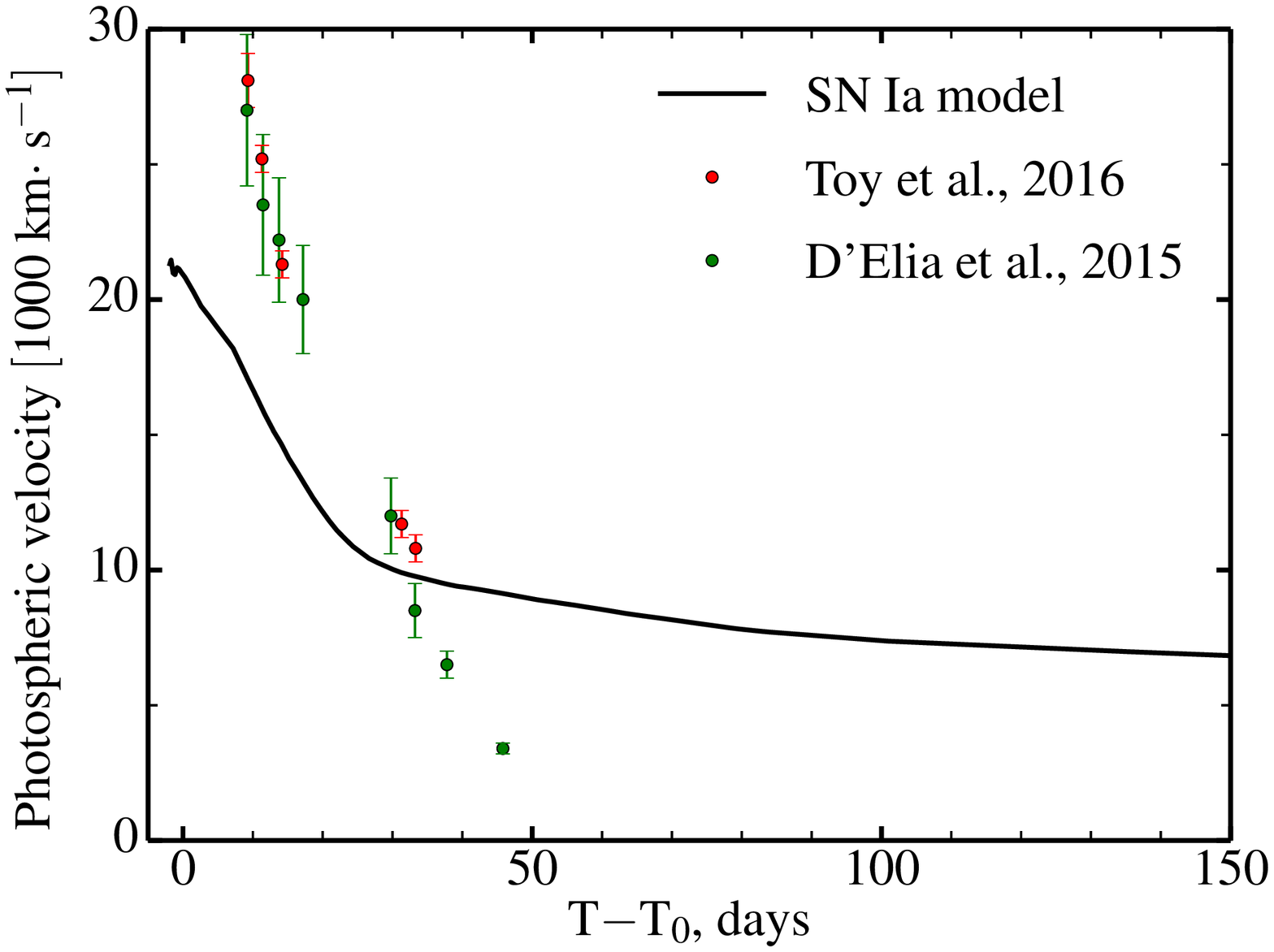} \\
a) & b) \\
\end{tabular}
\caption{a) The observed light curve of SN~2013dx (points) and the result of modelling with the SN~Ia parameters (solid lines). The quasibolometric light curve of the SN in AB photometric system obtained as a sum
of the fluxes in $ugriz$ filters, is marked as $\textrm{bol}_{ugriz}$. b) Corresponding photospheric velocities: observed (points) and obtained from the model (solid line).}
\label{SN1a}
\end{minipage}
\end{figure*}

\section{Discussion}
\label{sec:discussion}

\subsection{The nature of the extended emission of GRB~130702A}
\label{sec:disc-gamma}

In the Section~\ref{sec:gamma} we mentioned the extended emission of the GRB ~130702A observed in gamma-rays, and now we would like to discuss its possible nature.

The main component in
SPI-ACS data has a smooth shape which can be fitted with an exponential function
\citep{norris}. The extended emission in
both GBM and SPI-ACS data looks like a plateau with a cut-off at about 650 s after
the trigger. This component could be connected with a rapidly spinning magnetar
activity, which was initially formed during a core collapse stage. It is clear that the mass of
a compact remnant $6~\rm M_{\sun}$ (see Section~\ref{sec:modelling}) is more than Oppenheimer-Volkoff limit, 
and the cut-off may be a manifestation of a delayed collapse into a black hole of a spun-down
magnetar \citep{vietri-stella,lyons}, which is already proposed for the explanation of the extended emission
phase of some GRBs \citep[e.g.,][]{130831}. 

Alternatively, the source of a prompt 
emission can be a neutrino heated jet \citep{macfadyen}. Later, if an accretion rate 
is high enough, an activation of Blandford-Znajek jet \citep{komis-barkov} can be a source 
of a high energy extended emission \citep{barkov-poz}. An emission cut-off could be 
explained by a rapid decrease of the accretion rate. The latter scenario is less plausible 
because of high radius of a pre-supernova star and consequently large size of an accretion disc.
Detailed analysis of the extended emission of GRB 130702A will be presented elsewhere (Minaev et al., in preparation).

\subsection{The Amati relation}

Assuming a power-law model to be valid down to 1 keV (see Section~\ref{sec:obsgamma}), we
found the fluence of the burst to be $(1.31 \pm 0.08) \times 10^{-5}$ erg~cm$^{-2}$ in
the energy range 1--10000 keV. Assuming $z = 0.145$ and a standard
cosmology model with $H_{0} = 69.6$ km~s$^{-1}$~Mpc$^{-1}$, $\Omega_{M} = 0.286$,
$\Omega_{\Lambda} = 0.714$~\citep{Wright,bennett2014}, the isotropic energy release in 1--10000 keV is $E_{\rm iso,\gamma} = (6.6 \pm 0.4) \times 10^{50}$ erg in a GRB rest frame.

Almost all long GRBs follow $E_{\rm peak}\times(1+z) - E_{\rm iso}$ relation~\citep{amati}. If we assume that GRB~130702A also follows the Amati relation, then its $E_{\rm peak}\times(1+z)$ parameter value should be within 7--50 keV ($\pm$~2-$\sigma $ coincidence level). However, the energy spectrum of GRB 130702A is well fitted by a simple power-law model with index of $\gamma = -1.78 \pm 0.02$ and does not show presence of spectral break in energy range of 10--15000 keV. The value of the spectral index is typical for $\beta$ of Band function~ \citep{band93} and may indicate that $E_{\rm peak}$ is less than lower limit of analysed energy range: median values of spectral parameters obtained for GBM bursts see in tables 3-4 in~\citep{Gruber2014}. Under this assumption we estimate upper limit of $E_{\rm peak}$ to be approximately 10 keV. In this case GRB~130702A will follow the Amati relation. Similar result was also obtained by~\citet{amati-gcn}.

\subsection{Late X-ray light curve}
\label{subsec:xray}

In Section~\ref{sec:xrt} we have fitted the X-ray light curve of GRB~130702A with a BPL. There is an enhancement of the flux in the residuals plot (see lower panel of Fig.~\ref{xrt}). The flux of the enhancement, integrated in the time interval of $5-15$ d after the trigger, is $F^+ = (2.4 \pm 0.6) \times 10^{-12}$ erg cm$^{-2}$ s$^{-1}$, which is equivalent to the luminosity $L^+_{\rm 4\pi} = (1.4 \pm 0.3) \times 10^{43}$ erg s$^{-1}$. Here we use the luminosity distance $D_L = 692$ Mpc corresponding to $z = 0.145$. We can not distinguish weather it is related to a flare activity of the GRB afterglow or to an X-ray radiation from the SN component. If this enhancement is related to the SN component, the corresponding luminosity is 4 orders higher than the X-ray luminosity of the well known SN~1993J \citep[SN IIb,][]{1993J}. However, there are some GRB-SNe discussed in the literature \citep[e.g., ][]{li-pun}, where X-ray luminosities of GRB-SNe events are comparable with our estimates.

\subsection{SN Energy budget}

In the Section \ref{sec:modelling} we derived a value of the total energy of the SN explosion $E_{\rm oburst} = 3.5 \times 10^{52}$ erg and the bolometric energy of the SN radiation from far IR to extreme UV $E_{\rm bol} = 3.1 \times 10^{49}$ erg. We would like to estimate the radiative efficiency of the SN $\eta = E_{\rm rad} / E_{\rm oburst}$, where $E_{\rm rad}$ is an amount of the energy released in the radiation in both optical range and X-rays, $E_{\rm rad} = E_{\rm bol} + E_{X, \rm SN}$. Suggesting that the enhancement of the X-ray flux in the period of $5-15$ d after the trigger is related to the SN activity (see Section~\ref{subsec:xray}), we estimate the energy of the SN released in the X-ray domain as $E_{X, \rm SN} = 1.3 \times 10^{49}$ erg. Compiling all these values the radiative efficiency of the SN explosion is $\eta > 0.1$ per cent. Otherwise, if we suggest that the SN X-ray emission is negligible, then we obtain a conservative lower limit of the efficiency $\eta \geq 0.08$ per cent.

\subsection{Comparison with other GRB-SNe}

In our model we obtained the pre-supernova star mass $M = 25~\rm M_{\sun}$ and the mass of the resulting compact remnant $M_{\rm CR} = 6~\rm M_{\sun}$. Hence, the total mass of the ejecta is $M_{\rm ej} = 19~\rm M_{\sun}$.
The large ejecta mass and the possible presence of clumps of matter around the progenitor (bump in red filters) are consistent with the explosion of a rather massive star and suggest that the progenitor of SN 2013dx was a massive Wolf-Rayet star, whose strong winds drove the stripping of the outer layers. The higher ejecta mass is predicted for some Type Ic supernovae from a massive progenitor~\citep[e.g., SN~2011bm, ][]{Valenti2012}. The final mass of black hole is compatible with current observations and theoretical predictions \citep[see Figure~20 of~][]{Sukhbold2016}. However, the ejecta mass is greater than that obtained by D15 and T16, and also greater than ejecta masses of other GRB-SNe, collected by T16 and by \citet[][see Table~1 there]{cano16}. On the contrary, the mass of synthesized nickel $M_{\rm^{56}Ni}$ is lower than that averaged by all GRB-SNe and barely falls in 1-$\sigma$ interval around the average value. Concerning the total energy of the explosion $E_{\rm oburst}$, it is greater than the average value and falls into the top third of all GRB-SNe total energy list.

\citet{SNwhitedwarf} proposed a model of supernova-like explosion induced by gamma-ray burst in a binary system. The ejecta of a GRB exposes a white dwarf companion and initiates Type Ia supernova explosion. It would lead to SN~Ia phenomenon associated  with gamma-ray burst. Using \textsc{stella} code we modelled the light curves of SN~2013dx with SN~Ia model ($M_{\rm^{56}Ni}=0.35~\rm M_\odot$). The resulting light curves fit rather well the data (see Fig. \ref{SN1a}), but the photospheric velocities calculated using this model do not correspond to those measured spectroscopically by D15 and T16. Therefore, the SN~Ia model should be discarded.

In T16 the main parameters of the SN were obtained using the SN~1998bw as a template. But assuming those parameters as input for the modelling we could not obtain a good agreement (see Section~\ref{sec:modelling} and Fig.~\ref{lc-toy}). The parameters reported by T16 do not allow us to reproduce the observations by the model. Thereby we suggest, that the light curve of SN~1998bw is not an universal template to describe GRB associated supernovae. In the case of SN~2013dx the template of SN~2003dh, used in D15, gives better input parameters for the light curve modelling.

\subsection{Remaining questions}
\label{sec:remain}
The resulting model is the best from all of the models considered, still there remain some problems to be resolved.

\begin{enumerate}
\item There is a problem with simultaneous fitting of observational data in all filters. The time position of the maximum in the modelled and observational light curves does not coincide in every observational range, especially in blue filters. There is an option to vary the start time of the outburst $T_0$, but for good agreement between the model and the data one need to choose different $T_0$ for blue and red filters. But this would break the self-consistence of the model. This problem may be accounted for by the asymmetry of the outburst. 
\item Despite the modelled quasibolometric light curve fits well observational data, the multicolour model does not fit well the peak of the SN in blue filters. The discrepancy between observational points in $B$ and $u$ filters and resulting modelled light curves may be explained by additional absorption along the line of sight which is not included in the host extinction. In $u$ filter near the observed maximum the difference between the observations and the model is $A_u \sim 1^m$ corresponds to $A_V \sim 0\fm3-0\fm4$ depending on the adopted extinction law. Assuming the wind-like medium and the normalization density parameter $A_{\ast}$ from \citet{singer_paper} (see Sec.~\ref{sec:xrt}) we estimated the hydrogen column density $N_H$ in the host galaxy of the source integrating the profile from the radius of the pre-supernova star $R = 100~\rm R_{\sun}$ to infinity. We obtained $N_H = 3.7 \times 10^{19}$ cm$^{-2}$. Assuming the relation between $N_H$ and $A_V$ proposed by \citet{nh-av} for the Milky Way we estimate the extinction along the line of sight to be $A_{V}^{\rm LOS} \sim 0.02^m$. This estimation does not allow us to explain the significant difference between the modelled and the observed light curves in blue filters by some extra extinction along the line of sight or in the host galaxy.
\item The resulting model does not describe the secondary bump observed during the SN decay phase clearly visible in red filters (see Fig.\ref{lightcurve}). The bump may be connected not with the SN, but with the afterglow. In this case it may be explained as an interaction of the ejecta with some inhomogeneities in the surrounding interstellar medium, e.g., dense interstellar clouds. The theoretical model when the prompt $\gamma$-ray emission re-radiates on some dense shell or cloud in the medium surround the burst progenitor and forms bumps on the light curve was proposed by \citet{bk-tim} and there are a few papers with numerical simulations \citep{barkov-bk,postnov,badjin}. The model of the thick shell or cloud predicts the chromatic behaviour of the bump, i.e., the bluer the filter the earlier the bump, and a relative brightness in bluer filters should be more pronounced than in red ones. Our observations clearly support the prediction about chromatic behaviour, but the flux of the observed bump does not agree with the prediction: on the light curves (Fig.~\ref{lightcurve}a) we clearly see the bump in $z$ filter, while there is no bump in $g$ filter. However, it might be explained by a sparse sampling of the light curve in $g$ filter. Moreover, explanation of the bump requires specific configuration, in particular, there should be a compact dense cloud off-axis to observer, but not to far to be within the emission cone. The duration of the bump ($\tau_{\rm bump} \sim 10$ days) is explained in numerical simulations~\citep{badjin} with the \textsc{stella} code. 

Similar bumps present on the light curves of other Type Ib/c SNe, SN~2003dh associated with GRB~030329, but in this case it is difficult to distinguish the behaviour of the SN light curve from that of the extremely non-monotonic afterglow \citep[see, e.g., ][and references therein]{lipkin, cano16}. There is a possibility in the \textsc{stella} calculations to take into account additional interaction of the SN ejecta with the interstellar medium by surrounding the star by a shell (or super-wind) which could induce the bumps on the modelled light curve \citep{Sorokina2016}. We did not implement super-wind in current modelling. However, the bump has appeared in some other models in which $^{56}$Ni is put in the centre of explosion without mixing. The absence of the bump on the optimal resulting modelled light curve (Fig.~\ref{lightcurve}a) may be also connected to the simplicity of the assumptions, which underlie the model and the code itself.

It is unlikely, that the late bump is related to the central engine activity, and thus, to the afterglow evolution. This bump could be related to the re-radiation of the prompt gamma-ray emission by a clumpy circumburst medium, but in this case one could expect to observe several bumps from the interaction of gamma-rays with every clump around the source \citep[e. g., the late light curve of GRB~030329][]{lipkin}. The observed bump could be natural if the circumburst medium consisted of only one single clump. We can not definitely determine the bump's nature and we suggest that it is related to the SN. Similar bumps may emerge in the models with different $^{56}$Ni distribution inside a pre-supernova star \citep[][see also Fig.~\ref{sens}b,c,d]{moriya}.
\end{enumerate}

\section{Conclusions}
The light curve of SN~2013dx associated with GRB~130702A is the second well sampled GRB-SN after SN~1998bw. We collected all available optical data of this event: the multicolour light curves of GRB~130702A contain 330 data points in filters $uBgrRiz$ until 88 days after the burst start, more than 280 of them form the light curves of the associated supernova SN~2013dx. 40 of these points are published for the first time.

We presented the multicolour light curves of this SN, modelled with the code \textsc{stella}. In general, the model in filters as well as quasibolometric light curve is in a fairly good agreement with the observations. Moreover, the STELLA predictions of photospheric velocities fit well the ones obtained from spectra. The bolometric parameters of the supernova according to the model are: $M = 25~\rm M_{\sun}$, $M_{\rm CR} = 6~\rm M_{\sun}$, $E_{\rm oburst} = 3.5 \times 10^{52}$ erg, $R = 100~\rm R_{\sun}$, $M_{\rm^{56}Ni} = 0.2~\rm M_{\sun}$ and it is totally mixed inside the envelope; $M_{\rm O} = 16.6~\rm M_{\sun}$, $M_{\rm Si} = 1.2~\rm M_{\sun}$, and $M_{\rm Fe} = 1.2~\rm M_{\sun}$.

Disagreement between the modelled flux and the data increases from red to blue filter (from $g$ to $u$). Comparison between the model and the data at the peak of the light curve suggests an evidence for an additional line-of-sight extinction in the host galaxy, but the presence of this additional extinction is not confirmed by other methods of investigation, namely, by modelling the host SED and circumburst environment density. Instead, it might be related with a non-homogeneous mixing of $\rm^{56}$Ni.

A modelling of the SN light curve allowed us to estimate the conversion factor between the total energy of the SN outburst and the energy emitted by the SN as electromagnetic radiation to be 0.1 \%.

GRB~130702A is one more GRB with the extended emission detected in the gamma-ray light curve. Since the central engine is thought to power the extended emission, this might suggest that GRB 130702A's central engine is similar to those of GRB~111209A and GRB~130831A. 

\section*{acknowledgments}
We are grateful to M. Barkov, S. Grebenev, S. Moiseenko, B. Fain, and V. Loznikov for useful discussions. We are thankful for the anonymous referee for valuable remarks.
The work of A. A. V., A. S. P., P. Yu. M., and E. D. M. (data reduction, analysis, interpretation and comparison
with other GRB-SNe) was supported by Russian Science Foundation grant 15-12-30016. 
M.V.P. acknowledges support from RSCF grant no. 14-12-00146 for SN modelling with \textsc{stella} code.
This work of S. I. B. (development of \textsc{stella} code) was supported by Russian Science Foundation grant 14-12-00203.
Grant no. IZ73Z0\_152485 SCOPES Swiss National Science Foundation supports work of S. I. B. on production of radioactive elements in stellar explosions.
The work at Abastumani was supported by the Shota Rustaveli
National Science Foundation, Grant FR/379/6-300/14.
All observations at Maidanak astronomical observatory was supported by grant of "Committee for coordination science and technology 
development of Uzbekistan" (No.: F2-AS-F026). We gratefully acknowledge to observers B.~Khafizov and O.~Parmonov for support 
observations at Maidanak astronomical observatory. O.~A.~B. is thankful to the Matsumai International Foundation (MIF, Tokyo, Japan) 
(research fellowship ID No.: 15G24).
This work made use of data supplied by the UK Swift Science
Data Centre at the University of Leicester. This research has made
use of NASA's Astrophysics Data System.


\bibliographystyle{mn2e}

\label{lastpage}
\end{document}